\documentclass[aps,prd,amsmath,amsfonts,amssymb,eqsecnum,nofootinbib,
  floatfix,secnumarabic,preprintnumbers,superscriptaddress,nobalancelastpage,
  onecolumn,notitlepage,biblatex]{revtex4-1}
\usepackage{graphicx}
\usepackage{dcolumn}
\usepackage{bm}
\usepackage{hyperref,xcolor}
\usepackage[bottom]{footmisc}
\usepackage{fancybox}
\usepackage{cancel}
\usepackage{diagbox}
\usepackage{caption}
\usepackage{subcaption}
\usepackage{float}
\restylefloat{table}
\usepackage{dcolumn}   
\usepackage{bm}        
\usepackage{multirow}
\usepackage{blindtext}
\usepackage{enumitem} 
\usepackage{bm}    
\usepackage{adjustbox}
\def\h{h^0}

\def\hpm{H^{\pm}}

\def\wpm{W^{\pm}}
\def\wmp{W^{\mp}}

\begin{document}
	

\title{Charged Higgs Prospects In Extended Gauge Models}

\author{Baradhwaj Coleppa}
 \email{baradhwaj@iitgn.ac.in}
 \author{Gokul B. Krishna}
 \email{gokulb@iitgn.ac.in}
\author{Agnivo Sarkar}%
 \email{agnivo.sarkar@iitgn.ac.in}
\affiliation{IIT Gandhinagar, Palaj Campus, Gujarat 382355, India}%


\begin{abstract}

In this paper, we explore the collider phenomenology of the charged Higgs boson in the context of a generic Beyond Standard Model scenario with extended gauge and scalar sectors. In such scenarios, the charged Higgs boson can decay via the $W^{'}Z/ WZ^{'}$ channels. We formulate a search strategy for the $H^{\pm}$ in the channel $\sigma(g b \rightarrow H^{\pm}t)\mathcal{BR}(H^{\pm} \rightarrow W' Z)$ considering the interesting cascade decay chain $H^{\pm} \rightarrow W^{'} Z \rightarrow W^{\pm} Z Z$. We find that the charged Higgs can be discovered in final states with multiple hard leptons and/ or b-quarks which future LHC experiments with sufficiently large luminosity ($\mathcal{L} = 1000 fb^{-1}$ and above) can probe.   

\end{abstract}

\maketitle

\section{Introduction}

The Standard Model (SM) of particle physics has enjoyed tremendous success over the years and with the discovery of the Higgs boson at the LHC in 2012 \cite{Aad:2012tfa,CMS:2012qbp}, its particle spectrum is now firmly established. However, in parallel, a large body of experimental and theoretical work suggest that there has to lie some physics beyond the SM - generically dubbed BSM physics or New Physics (NP) - at the TeV scale, and the LHC experiments are currently working to uncover hints of BSM physics in a wide variety of final states. Construction of BSM avenues typically involves enlarging the scalar, gauge sectors, or the matter content of the SM - there are many BSM scenarios in which more than one of these three sectors is modified. Correspondingly, experiments in high energy physics look for the new, heavy particles in such theories by typically taking advantage of the couplings of these heavy particles to the SM sector. This is a reasonable approach as one expects that the discovery potential should be maximal in process of the form NP$\to$ SM, SM. However, given the non-observation of any BSM particle in the LHC experiments thus far, it behooves one to ask if there are other channels that one could probe that might uncover hints that traditional search channels might have missed. In this work, we undertake one such study for heavy charged Higgs bosons that appear in many scalar extensions of the SM.

Extending the scalar sector with new Higgs fields in doublet or other representations of the SM $SU(2)_L$ has a long history. Examples include the Two Higgs-Doublet Models (2HDMs)\cite{Branco:2011iw} and their many variants\cite{DeCurtis:2016tsm}\cite{Logan:2010ag}, the minimal supersymmetric Standard Model (MSSM)\cite{Drees:1988fc}\cite{Haber:1984rc}, and other singlet extensions of the SM. Many such models typically involve enlarging only the scalar part of the SM Lagrangian leaving the SM gauge group $SU(2)_L\times U(1)_Y$ as is. However, it is equally possible to consider scenarios wherein one extends the SM gauge group simultaneously involving additional Higgs fields thus invoking more complicated patterns of symmetry breaking. In models of these kinds, one expects, in addition to traditional searches, process of the form NP$\to$ SM, NP to be viable as well. If nature indeed chose to follow such a path, it would be interesting to explore the consequences of what an extended gauge model could mean for charged Higgs prospects at the LHC - this simple framework forms the motivation for the present work.

\section{The $H^{\pm}$ Boson Discovery Prospect at LHC}
\label{Sec:Pheno}
\subsection{Set-up and Current Limits}
\label{Sec:setup}
Here, we will undertake the LHC study of charged Higgs bosons in theories which also have an enlarged gauge symmetry. At the outset, we make no assumptions about the exact gauge group or, for that matter, the representations of the additional Higgs fields under the SM and the extra gauge groups. Our discussion in this section will be mostly model-independent with the following minimal set of assumptions:
\begin{itemize}
\item The BSM sector allows for the existence of heavy, charged Higgs bosons $\hpm$ in addition to the SM Higgs $\h$. There could (and in general, will) be other neutral and/or charged Higgs bosons, but our analysis in this section is blind to these specifics.
\item In addition, the model also allows for the presence of new, heavy charged and neutral gauge bosons ($W'$, $Z'$) that are \emph{fermiophobic}. This last assumption is a critical one as we would like to explore charged Higgs decay to these new gauge bosons, and will therefore require relatively light $W'$ and $Z'$ that are not already ruled out by direct searches.
\item There is a non-zero coupling between the heavy gauge bosons and the $H^\pm$ at tree level, i.e., the vertices of the type $H^\pm W'^{\mp}Z$ and $H^\pm W^{\mp}Z'$ exist and are non-vanishing.
\end{itemize}
Our goal here is to explore the discovery prospects of the charged Higgs in the channel $\hpm\to W'^{\pm},Z$. A simpler alternative would of course, be to simply look for the process $\hpm\to \wpm Z$ - the $\hpm\wmp Z$ is an interesting vertex whose phenomenology has been explored elsewhere \cite{Cen:2018okf}\cite{Adhikary:2020cli}. It has been shown \cite{Gunion:1989we} that for the Lagrangian term $\mathcal{L} = \xi H^{\pm}W^{\mp}_{\mu}Z_{\mu} + h.c.$, the coupling can be expressed in the general fashion
\begin{equation}
\xi^{2} = \frac{g^{2}}{4m^{2}_{W}}\left[ \sum_{i}Y^{2}[4T_i(T_i + 1) - Y_i^{2}]v_i^{2}\right] - \frac{1}{\rho^{2}} ,
\end{equation}
where $\rho = \frac{m^{2}_{W}}{m^{2}_{Z}\cos^{2}\theta_{W}}$, $T_i$ and $Y_i$ are the $T_3$ and hypercharge quantum numbers of the $i$th Higgs field and $v_i$ is its vacuum expectation value (vev). For the 2HDM, the above coupling vanishes at tree level  \cite{Gunion:1989we} - in fact, for any model with multiple Higgs fields \emph{all} in the doublet representation under the SM $SU(2)_L$, it can be shown that the $H^{\pm}W^{\mp}_{\mu}Z_{\mu}$ vertex is absent at tree level \cite{Johansen:1982qm}. On the other hand, the Georgi-Marchak model\cite{Logan:2015xpa}\cite{Degrande:2015xnm} for instance, where one introduces an additional Higgs triplet, contains in its scalar spectrum a $H^{\pm}$ which does indeed couple to $W^{\pm}$ and $Z$ boson at tree level but remains fermiophobic. It is evident that, depending on model construction the phenomenology associated with various BSM states will alter significantly. With that in mind, in this article we propose an alternative discovery prospect of the charged Higgs boson within the context of a BSM scenario with enlarged scalar \emph{and} gauge sectors.

\begin{figure}[h!]
\begin{center}
\includegraphics[scale=0.36]{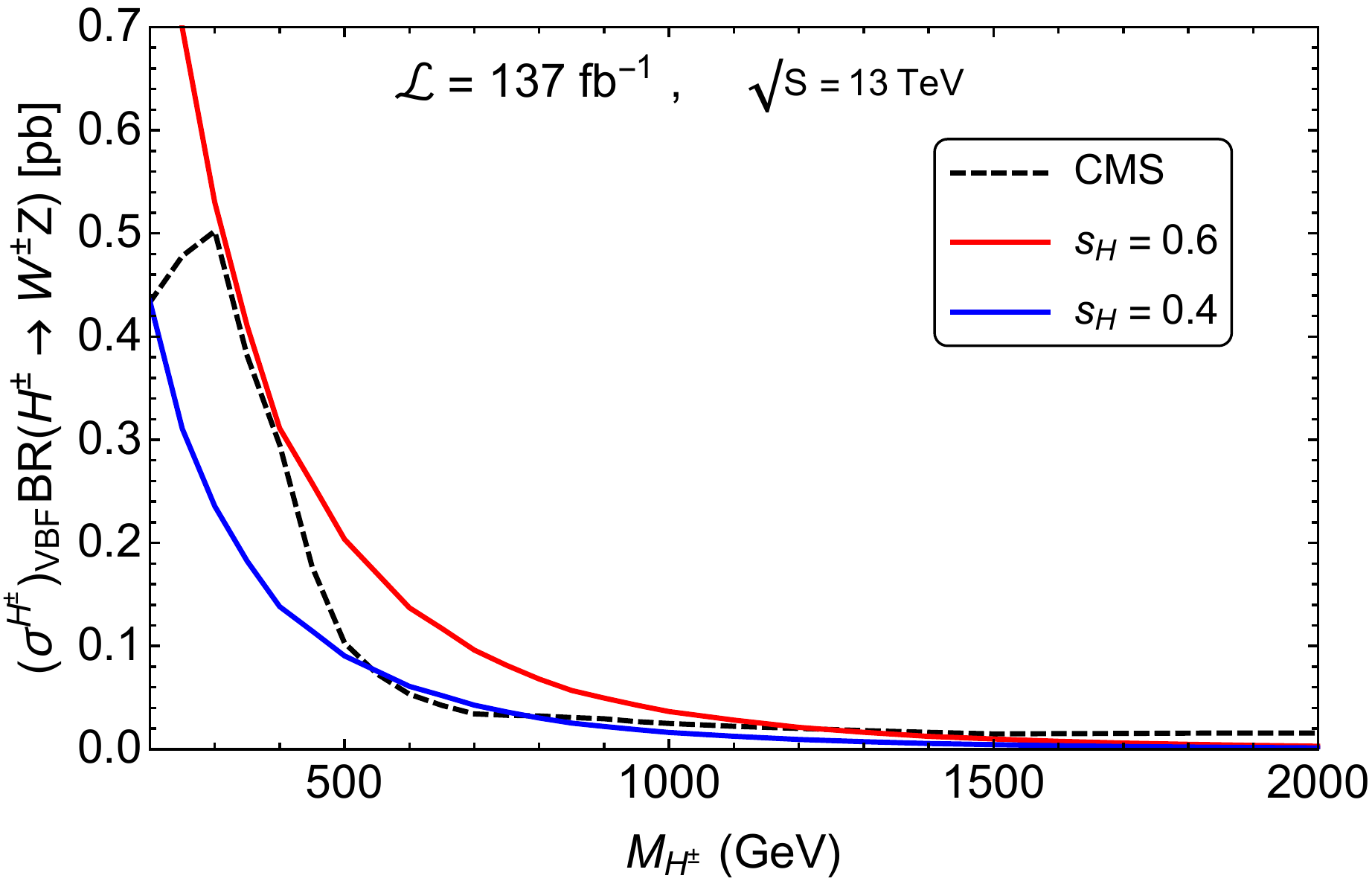}
\hspace{0.2in}
\includegraphics[scale=0.37]{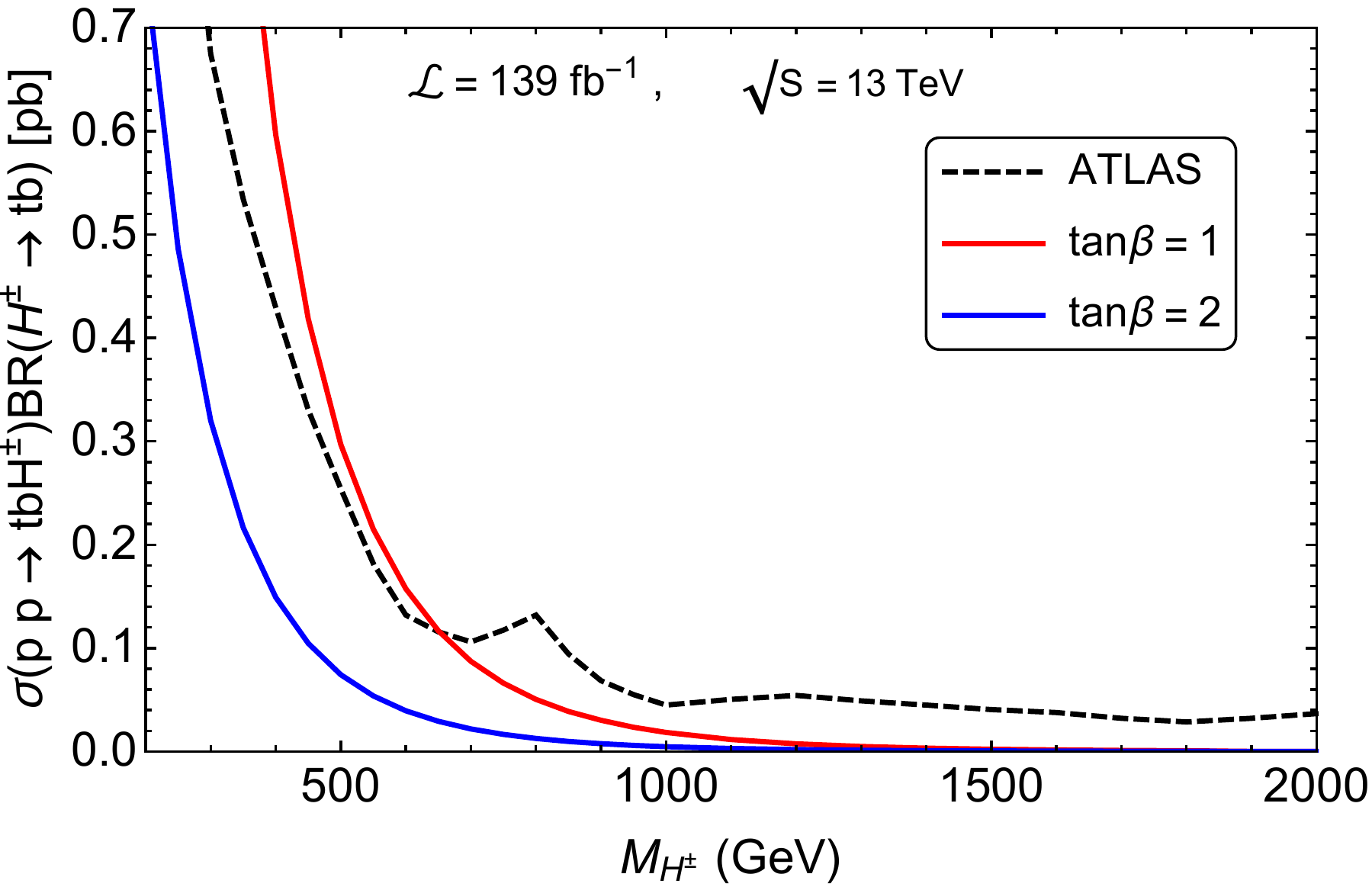}
\end{center}
\caption{The lower bound on the $H^{\pm}$ boson production cross-section obtained from the latest ATLAS measurements \cite{Aad:2021xzu} from the $tb$ mode production and decay (right) and the CMS measurements \cite{Sirunyan:2021lsf} from the vector boson fusion production mechanism (left). }
\label{fig:Exp_Hc}
\end{figure}

 Both the ATLAS \cite{Aad:2015typ,Aad:2015nfa} and CMS \cite{Sirunyan:2020hwv}\cite{Sirunyan:2021lsf} have looked for the $H^{\pm}$ primarily via two mechanisms: i) where the $H^{\pm}tb$ vertex comes into play in production or decay mode or both, and ii) where the $H^{\pm}$ is produced from the fusion of $W^{\pm}$ and $Z$. In Fig[\ref{fig:Exp_Hc}], we present the bound (obtained from ATLAS and CMS measurement) on the $H^{\pm}$ production cross-section for these two mechanism \cite{Aad:2021xzu}\cite{Sirunyan:2021lsf} - understandably, the limits are weaker in the latter case. Depending on the final state, the combined measurement by ATLAS has imposed upper limits on $\sigma \times $BR for a charged Higgs in various channels in the mass-range 180 GeV to 3 TeV. The CMS collaboration searches and results can be found in \cite{CMS:2016szv}. In Table \ref{tab:Hclimits}, we summarize the various search channels for the charged Higgs in the ATLAS and CMS experiments and the mass range probed in each case along with the range of cross-section limits. We close this section mentioning that in a realistic model, these limits will be relaxed somewhat as the branching ratio for each channel would typically be less than 100\% (experimental numbers are quoted assuming 100\% BR in the channel of interest). Indeed in models where the ``non-standard" decays of the charged Higgs are sizeable \cite{Coleppa:2014cca}\cite{Coleppa:2019cul}, the limits can be substantially weakened - we will illustrate this in Sec.~\ref{sec:model} in one context. In the section on LHC phenomenology that follows, our discussion will thus be completely general without assuming any restrictions on the charged Higgs mass or its couplings.

\begin{table}[h!]
	\centering
	\begin{tabular}{|c|c|c|c|c|c|}
	\hline
	Exp.    & Channel & $\sqrt{s}$ & $\mathcal{L}$ & Mass Range & $\sigma_{\text{upper}}$ (in pb) \\
	\hline
	ATLAS \cite{Aad:2021xzu} & $\sigma(p p \rightarrow H^{\pm}tb)\mathcal{BR}(H^{\pm} \rightarrow t b)$ & 13 TeV & 139 $fb^{-1}$ & 200 GeV - 2 TeV & 3.6  - 0.036  \\
	\hline
	CMS \cite{Sirunyan:2020hwv} & $\sigma(p p \rightarrow H^{\pm}t[b] + p p \rightarrow H^{\pm})\mathcal{BR}(H^{\pm} \rightarrow t b)$ & 13 TeV & 35.9 $fb^{-1}$ & 200 GeV - 3 TeV & 21.3  - 0.007 \\
	\hline
	ATLAS \cite{Aaboud:2018gjj} & $\sigma(p p \rightarrow H^{\pm}tb)\mathcal{BR}(H^{\pm} \rightarrow \tau^{\pm} \nu_{\tau})$ & 13 TeV & 36.1 $fb^{-1}$ & 90 GeV - 2 TeV & 4.2  - 2.5$\times 10^{-3}$  \\
	\hline
	ATLAS \cite{Aaboud:2016dig} & $\sigma(p p \rightarrow H^{\pm}t[b])\mathcal{BR}(H^{\pm} \rightarrow \tau^{\pm} \nu_{\tau})$ & 13 TeV & 3.2 $fb^{-1}$ & 200 GeV - 2 TeV & 1.9  - 15$\times 10^{-3}$  \\
	\hline
	CMS \cite{Khachatryan:2015qxa} & $\sigma(p p \rightarrow H^{\pm}t[b])\mathcal{BR}(H^{\pm} \rightarrow \tau^{\pm} \nu_{\tau})$ & 8 TeV & 19.7 $fb^{-1}$ & 200 GeV - 600 GeV & 2.0  - 0.13  \\
	\hline
	ATLAS \cite{Aaboud:2018ohp} & $\sigma(p p \rightarrow H^{\pm}jj \rightarrow W^{\pm}Zjj)$ & 13 TeV & 36.1 $fb^{-1}$ & 200 GeV - 900 GeV & 0.25 - 0.05  \\
	\hline
	CMS \cite{Sirunyan:2021lsf} & $\sigma(p p \rightarrow H^{\pm}jj)\mathcal{BR}(H^{\pm} \rightarrow W^{\pm}Z)$ & 13 TeV & 137 $fb^{-1}$ & 200 GeV - 3TeV & 0.43 - 0.02  \\
	\hline
	\end{tabular}
	\caption{The various search channels for a charged Higgs boson in the ATLAS and CMS experiments and the mass ranges they have probed.}
	\label{tab:Hclimits}
\end{table}

\subsection{LHC Analysis}
\label{Sec:lhc}

In this section, we analyze the feasibility for a 5$\sigma$ discovery of the charged Higgs concentrating on its decay to particles in the extended gauge sector. We restrict our attention to the associated production mode of the $H^\pm$ $g b \rightarrow H^{\pm} b$ and  perform the data simulation using the MADGRAPH5aMC@NLO event generator \cite{Alwall:2014hca} (working in the five flavor scheme). The SM backgrounds which are used for this study are generated via the in-built SM model file in the MADGRAPH repository. The parton level simulation from MADGRAPH is passed on to PYTHIA 6 \cite{Sjostrand:2006za} for jet showering and hadronization followed by detector level simulation in DELPHES 3 \cite{deFavereau:2013fsa}. The BSM model file used in the next section has been designed using FEYNRULES \cite{Christensen:2009jx,Alloul:2013bka}. 

The primary process we are interested in is  $p p \rightarrow H^{\pm} \bar{t} \rightarrow W^{'\pm} Z \bar{t}$. Since we assume that the $W'$ does not couple to fermions, we are left with the choice of considering the two decays\footnote{Given that we are also dealing with an extended scalar sector, it is equally possible to consider $W'\to W H$, with the $H$ being a heavy neutral Higgs. Since $H\to b\bar{b}$ is also admissible, the final state here can be considered within this context as well.} $W^{'\pm} \rightarrow W^{\pm} Z$ and $W^{'\pm} \rightarrow W^{\pm} h$. Accordingly, we look at three different signals depending on the subsequent decays of the various gauge bosons. These are
\begin{enumerate}[label=\emph{\alph*})]
\item Signal 1: $p p \rightarrow H^{\pm} \bar{t} \rightarrow W^{'\pm} Z \bar{t} \,(W^{'\pm} \rightarrow W^{\pm} Z) \rightarrow W^{\pm} W^{\mp} Z Z b \rightarrow 4j + 4\ell + b $,
\item Signal 2: $ p p \rightarrow H^{\pm} \bar{t} \rightarrow W^{'\pm} Z \bar{t} \, (W^{'\pm} \rightarrow W^{\pm} h) \rightarrow W^{\pm} W^{\mp} Z h \bar{b} \rightarrow 2j + 3\ell + 3b + \cancel{E}_{T} $, and
\item Signal 3: $p p \rightarrow H^{\pm} \bar{t} \rightarrow W^{'\pm} Z \bar{t}\, (W^{'\pm} \rightarrow W^{\pm} Z) \rightarrow W^{\pm} W^{\mp} Z Z \bar{b} \rightarrow 4j + 2\ell + 3b  $
\end{enumerate}

In choosing the various final states, our consideration has been to both find a channel that would aid a complete reconstruction of the $H^\pm$ while also minimizing the SM background to the extent possible. This has led us to look at final states that have a good hadronic component with also leptons and missing energy (usually the presence of the latter will lead to one or more electroweak vertex in the corresponding SM Feynman diagrams leading to lower cross-sections). Considering the multijet final states, the major experimental search challenges will arise from SM processes like $t\bar{t}$+jets, $VV$+jets and subdominant contributions from $VVV+$jets etc. In addition, we have also considered $t\bar{t}h$+jets , $Vh+$jets, $t\bar{t}V$ and $VVh+$jets \footnote{In all these backgrounds, ``$+$jets" includes events with $+0$, $+1$, and $+2$ jets.} as part of the entire SM background.

We employ the following set of basic identification cuts at the time of simulation that would eliminate events with low $p_T$ jets and leptons:
\begin{equation}
	p_{Tj} > 20 ~\text{GeV},~~~ p_{T\ell} > 10~ \text{GeV},~~~ |\eta_j|~ \leq ~5,~~~  \text{and} ~~~|\eta_\ell|~ \leq ~2.5.
\end{equation}
We have chosen a wider window for the pseudorapidity for jets as compared to the leptons to ensure that we do not lose many signal events. Further, we demand that all pairs of particles are optimally separated: 
\begin{equation}
\Delta R_{jj} = \Delta R_{bb} = \Delta R_{jl} =  \Delta R_{bj} = 0.4. 	
\end{equation}

Within this basic framework, we now move on to the task of optimizing the discovery process of the charged Higgs by designing kinematic cuts for the three different final states given above. No heavy $W',Z'$ has been discovered at the LHC thus far and so, beyond the assumption of the existence of a heavy gauge boson with a specific mass, we do not use any particular attributes of it to construct our cuts. Specifically, we do not impose an invariant mass cut around the $W'$ mass to filter out the SM background.  We reiterate here that what follows is a purely background analysis - the signal is represented by a fiducial cross-section and included here to ensure that the cuts do not affect it too much. We discuss the implications of our findings on a toy model Lagrangian in Sec.~\ref{sec:model}.

\subsubsection{Signal I}

We begin our discussion of the process  $p p \rightarrow H^{\pm} \bar{t} \rightarrow W^{'\pm} Z t \,(W^{'\pm} \rightarrow W^{\pm} Z) \rightarrow W^{\pm} W^{\mp} Z Z b \rightarrow 4j + 4\ell + b $ choosing as a benchmark point  $m_{H^{\pm}}$ = 500 GeV and $m_{W^{'}}$ = 350 GeV (hereafter dubbed \textbf{BP1}). In Table~\ref{tab:Sig1BP1}, we present the cut flowchart that minimizes the SM background effectively for this particular choice. While stronger $p_T$ cuts beyond the identification step would prove to be reasonably useful, we find that for this signal a cut on the hadronic transverse energy proves efficacious. It can be seen in particular that the $t\bar{t}+$ jets background goes down by more than 50\% once we demand $H_T \geq 400~ \text{GeV}$.  We then move to a set of self-evident identification cuts: $N_j \geq 4 $, $N_l \geq 4$  as we are dealing with a multi-jet, multi-lepton final state. While this certainly helps us get rid of a major chunk of background, it can be seen that it does so at the cost of signal reduction - this is to be expected given the current lepton identification percentages at the LHC. To deal with the $t\bar{t}$ events that remain after this step (as there are still events with semi-leptonic decays of the top quark that can survive these identification cuts), we demand a lower bound on $\cancel{E}_T$ for the surviving events. Finally, we reconstruct the harder of the two $Z$ bosons from the more energetic lepton pair (this is from the $H^{\pm} \to W^{'\pm} Z$ part of the process. While strictly speaking this step is superfluous given ultimately we want to reconstruct the $H^\pm$ itself, it is included here for its effect in bringing down the vector boson dominated part of the background. 

\begin{figure}[h!]
\includegraphics[scale=0.35]{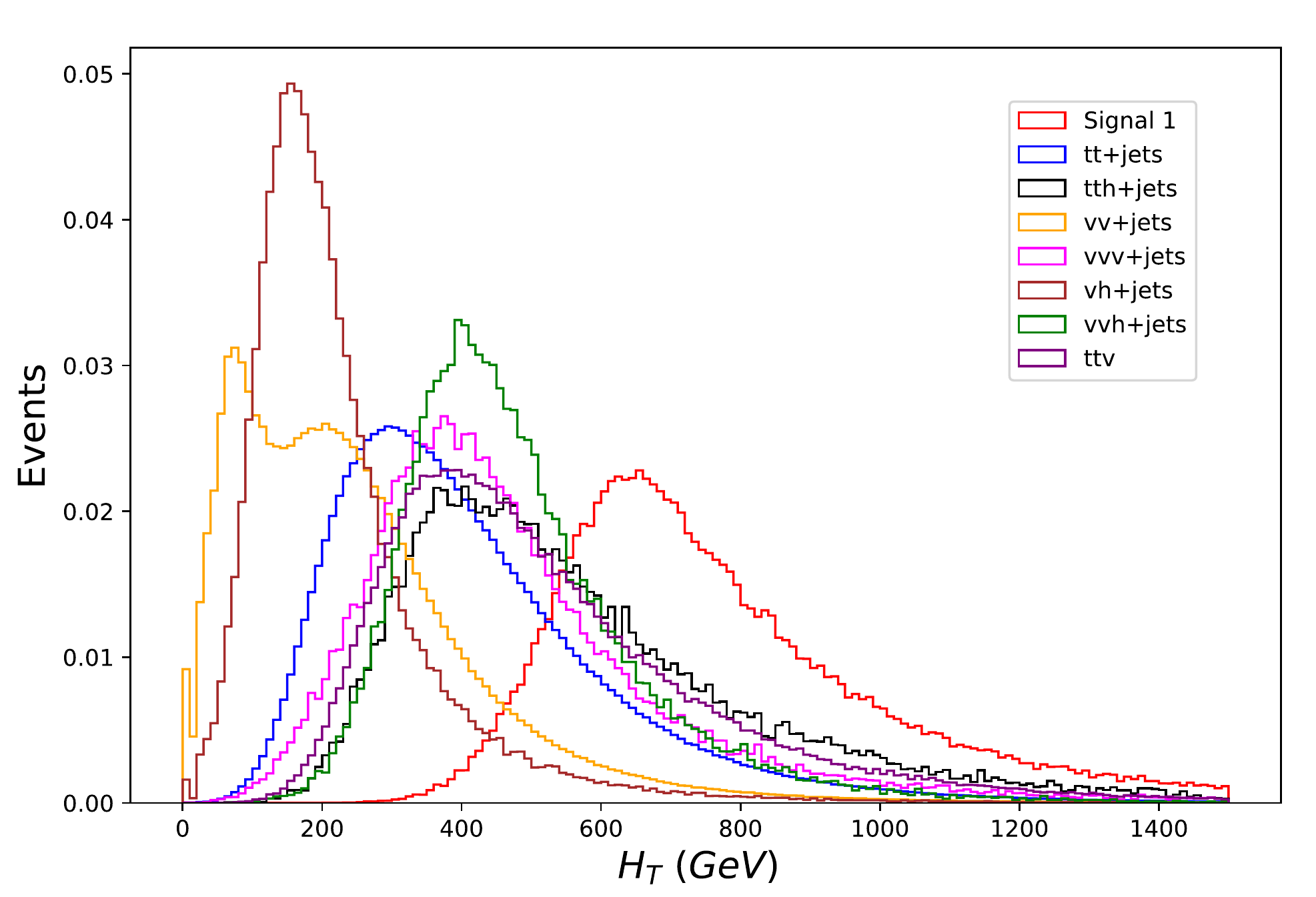}
\hspace{0.01in}
\includegraphics[scale=0.35]{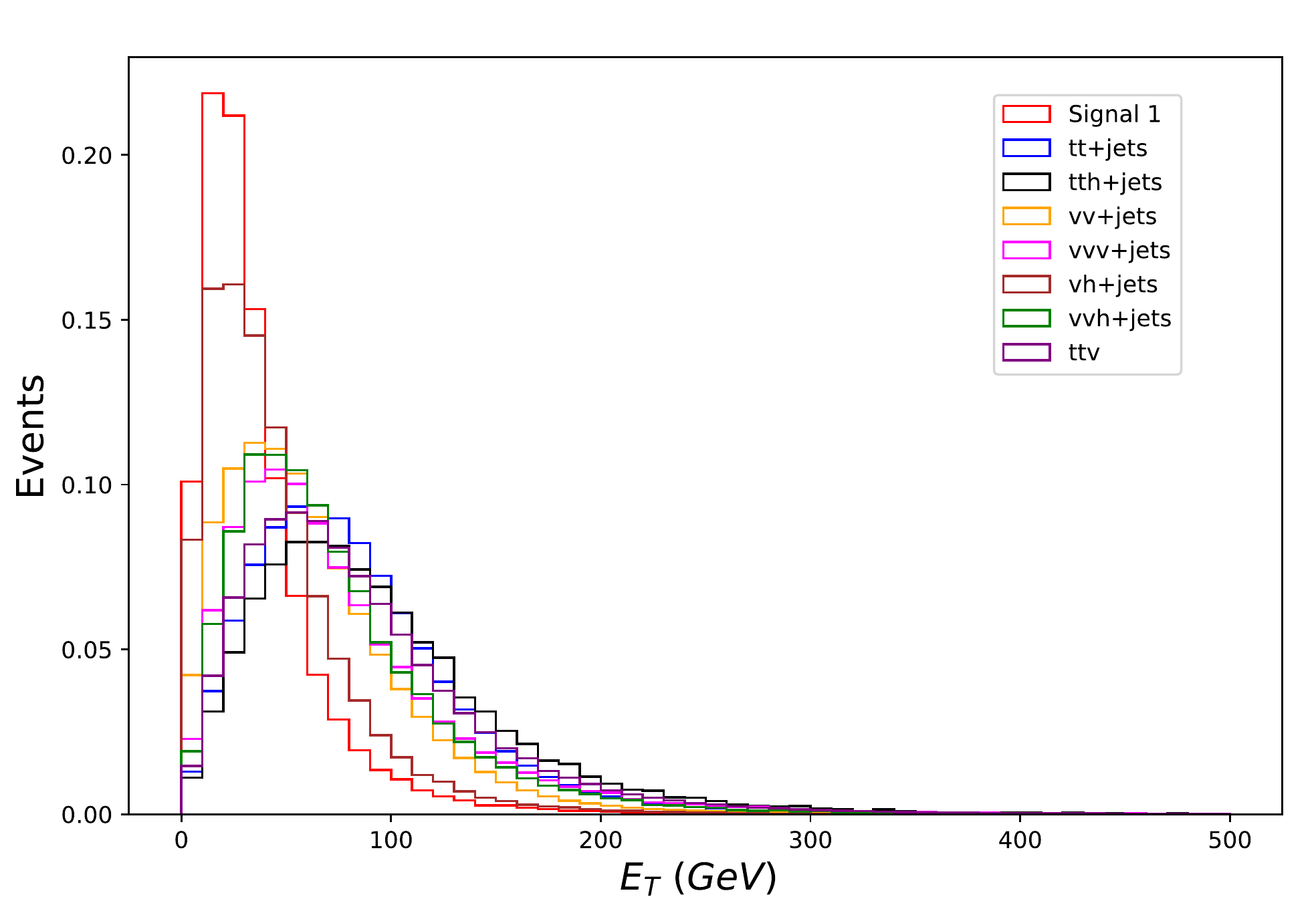}
\caption{The ${H}_T$ and $\cancel{E}_{T}$ distribution for both signal and SM backgrounds with $m_{H^{\pm}}$ = 500 GeV and $m_{W^{'}}$ = 350 GeV for Signal 1.}
\label{fig:S1_HT}
\end{figure}

\begin{figure}[h!]
\includegraphics[scale=0.35]{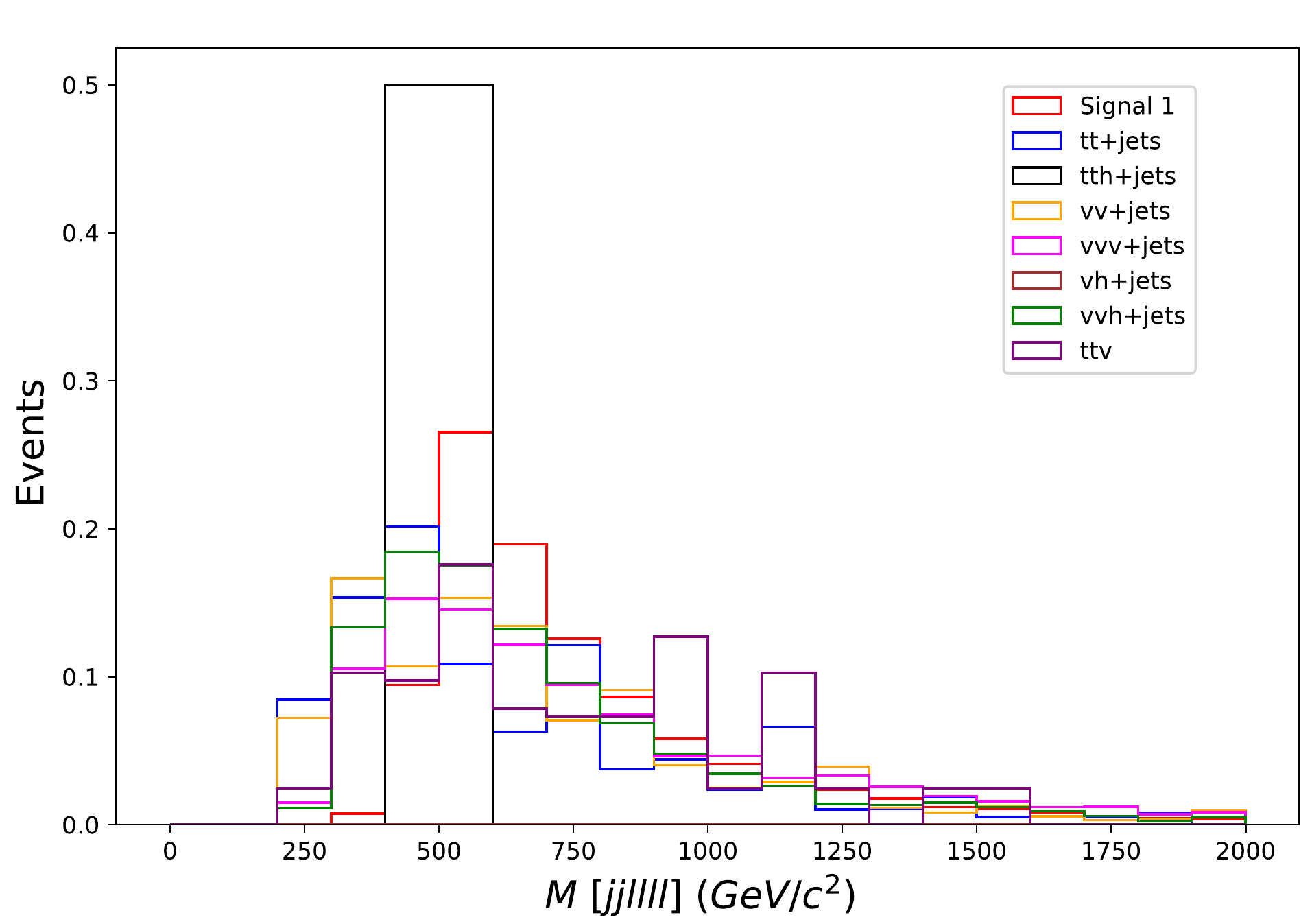}
\hspace{0.01in}
\includegraphics[scale=0.35]{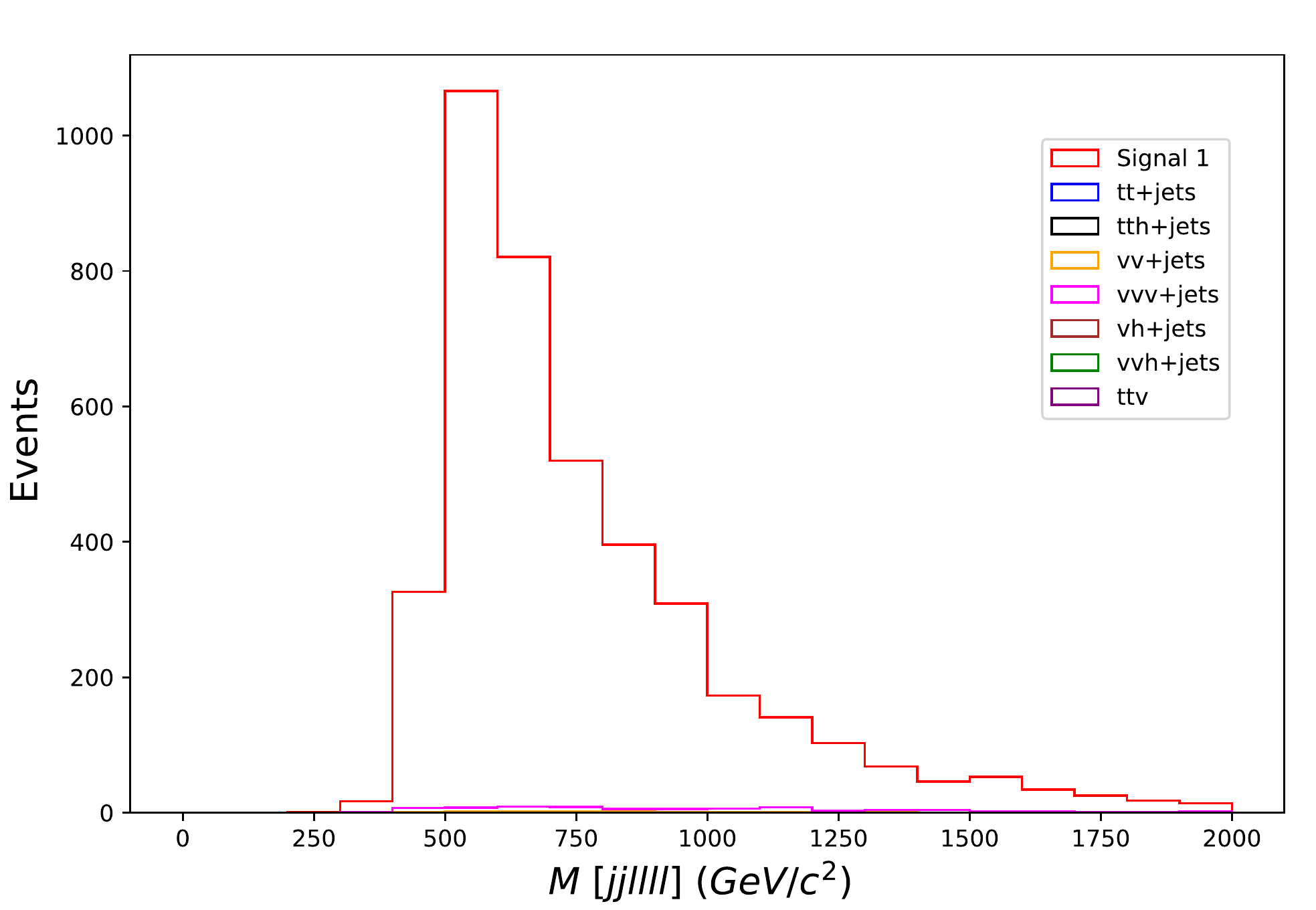}
\caption{The charged Higgs invariant mass distribution before and after implementing the cut mentioned in Table[\ref{tab:Sig1BP1}] for Signal 1.}
\label{fig:S1_invariant}
\end{figure}

The reason for choosing the particular numbers in Table~\ref{tab:Sig1BP1} for the $H_T$ and $\cancel{E}_T$ cuts can be inferred from Fig.~\ref{fig:S1_HT}. It can be seen that for the specific benchmark point chosen, the signal events (red curve in the plots) are peaked for larger $H_T$ ($>$ 400 GeV) and smaller $\cancel{E}_T$ ($<$50 GeV), regions where the SM background events are (mostly) minimal. While a stronger $H_T$ cut would undoubtedly help in reducing $t\bar{t}$ even more, we have stuck with a more moderate cut so as to not affect the signal too much. In Fig.~\ref{fig:S1_invariant}, we display the invariant mass distribution $m_{2j+4\ell}$ for the signal and background before and after implementing the kinematic cuts outlined in Table~\ref{tab:Sig1BP1} (here the two leading jets in the final state are chosen to reconstruct the $H^\pm$). Based on the distribution, we conclude that a slightly asymmetric cut of $400~\textrm{GeV}\leq m_{2j+4\ell}\leq 800~\textrm{GeV}$ would be quite effective in teasing out the signal in this particular channel.

\begin{table}[h!]
	\centering
	\begin{tabular}{|c|c|c|c|c|c|c|c|c|c|}
		\hline
		Cuts & Signal         & $t\bar{t}$+jets            & $VV$+jets             & $t\bar{t}h$+jets &$VVV+$jets   & $Vh$+jets & $VVh$+jets & $t\bar{t}V$ \\ \hline
		     &  100000      & 4350000         &  2300000   &  100000 & 500000 & 190000 & 270000 & 200000 \\ \hline 
		$H_{T} \geq$ 400 GeV & 99011& 1912953 & 426732 & 72508 &  285408 & 17125 & 176719 & 130377 \\ \hline 
		$N_{j} \geq$ 4 & 86480  & 1261763 & 197852 & 37199 & 143580 & 6145 & 86785 & 76641 \\ \hline
		$N_{\ell} \geq$ 4 & 12706  & 47 & 295 & 0 & 737 & 0  & 1 & 17\\ \hline
		$ \cancel{E}_T  \leq $ 50 GeV & 10332 & 6 & 257 & 0 & 300 & 0 & 1 & 5\\ \hline
		70 GeV $\leq M_{\ell\bar{\ell}} \leq$ 120 GeV & 4183 & 0 & 71 & 0 & 83 & 0 & 0 & 1 \\ \hline
	\end{tabular}
	\caption{Cut flow chart for the $4j + 4l + b$  channel with the signal corresponding to $m_{H^{\pm}}$ = 500 GeV and $m_{W^{'}}$ = 350 GeV.}
	\label{tab:Sig1BP1}
\end{table}

While the benchmark point chosen above certainly seems conducive to discovery in the $4j + 4\ell + b $ channel, it is instructive to ask how the discovery prospects look like for a case where there is a larger mass splitting between the $H^\pm$ and the $W'^\pm$. In Table~\ref{tab:Sig1BP2}, we present the cut flowchart for the benchmark point $m_{H^{\pm}} = 700$ GeV and $m_{W^{'}} = $350 GeV (hereafter called \textbf{BP2}).  Comparing with Table~\ref{tab:Sig1BP1}, we see a couple of important differences: since we are dealing with a more massive charged Higgs, we have chosen a stronger $H_T$ cut in this case. Accordingly, there is a larger reduction in the $t\bar{t}$ (and other) backgrounds. In addition, while it was not necessary in the previous benchmark point analysis, in this case we have put in a $p_T$ cut on the leading lepton to aid suppression of the SM background more effectively. While it can be seen that one can achieve a good $S/\sqrt{B}$ in this case as well, we need also to consider that in a realistic model scenario the production cross-section for a heavier $H^\pm$ will be significantly lower.

\begin{table}[h!]
	\centering
	\begin{tabular}{|c|c|c|c|c|c|c|c|c|c|}
		\hline
		Cuts & Signal  & $t\bar{t}$+jets            & $VV$+jets             & $t\bar{t}h$+jets &$VVV+$jets   & $Vh$+jets & $VVh$+jets & $t\bar{t}V$\\ \hline
		-      &  100000 & 4350000         &  2300000   &  100000 & 500000 & 190000 & 270000 & 200000\\ \hline 
		$H_{T} \geq$ 600 GeV & 95246 & 668744 & 117346 & 35957 &  106306 & 4880 & 51502 & 57907\\ \hline
		$N_{j} \geq$ 4 & 82523& 514346 & 68222  & 22548 & 57965 & 2264 & 24477 & 38144  \\ \hline
		$N_{\ell} \geq$ 4 & 16826& 35 & 160  & 0 & 497 & 0 & 0 & 11 \\ \hline
		$ \cancel{E}_T  \leq $ 50 GeV & 13099  & 5 & 137 & 0 & 727 & 0  & 0 & 2\\ \hline
		$p_T ({\ell}) \geq$ 100 & 12655 & 5 & 126 & 0 & 170 & 0 & 0 & 2  \\ \hline
		575 GeV $\leq M_{jjllll} \leq$ 1050 GeV & 8618 & 4 & 58 & 0 & 66 & 0 & 0 & 1\\ \hline
	\end{tabular}
	\caption{Cut-flow chart for the signal $4j + 4\ell + b$ channel with the signal corresponding to $m_{H^{\pm}} = 700$ GeV and $m_{W^{'}} = $350 GeV.}
	\label{tab:Sig1BP2}
\end{table}

\subsubsection{Signal 2}

\begin{figure}[h!]
\includegraphics[scale=0.35]{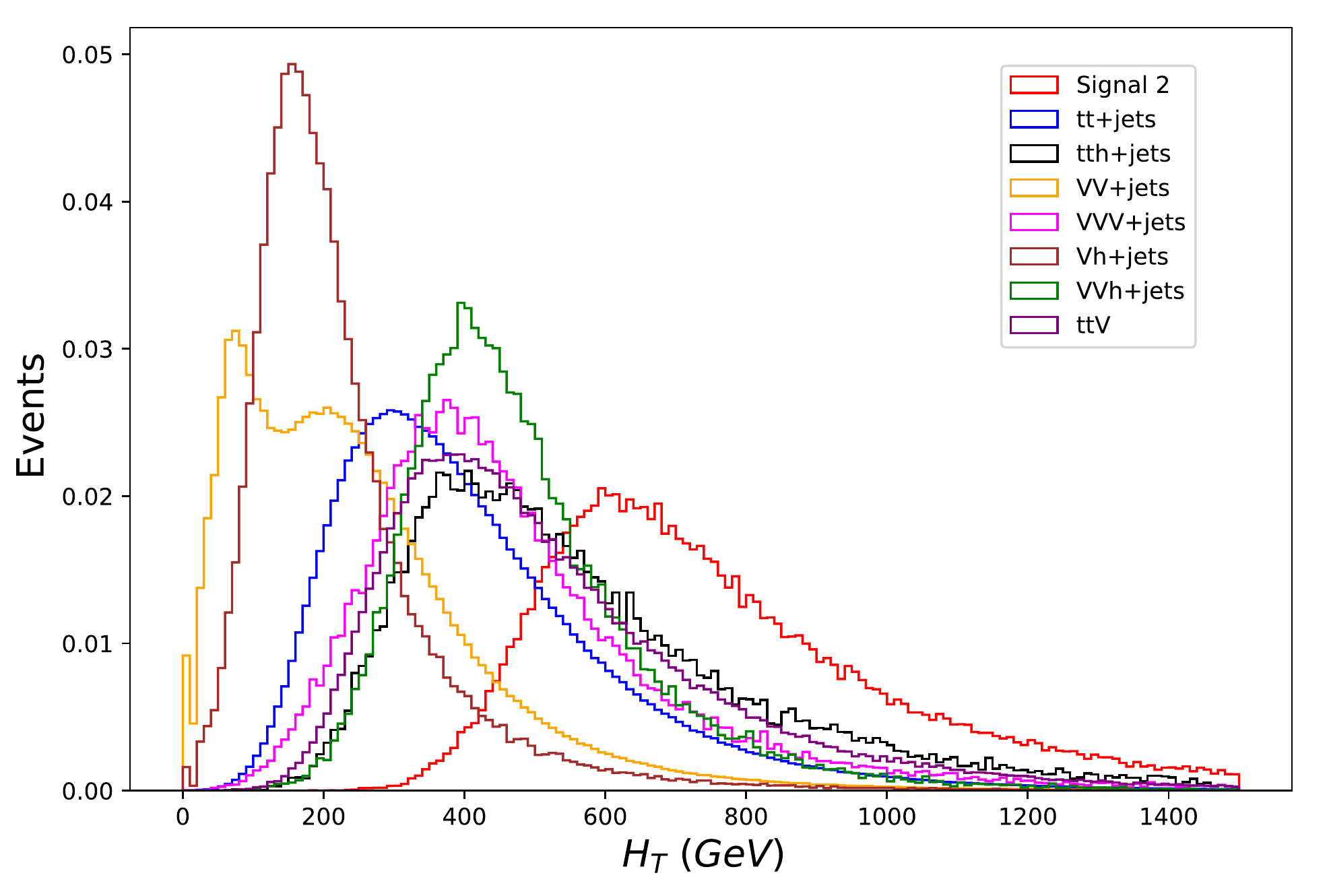}
\hspace{0.01in}
\includegraphics[scale=0.35]{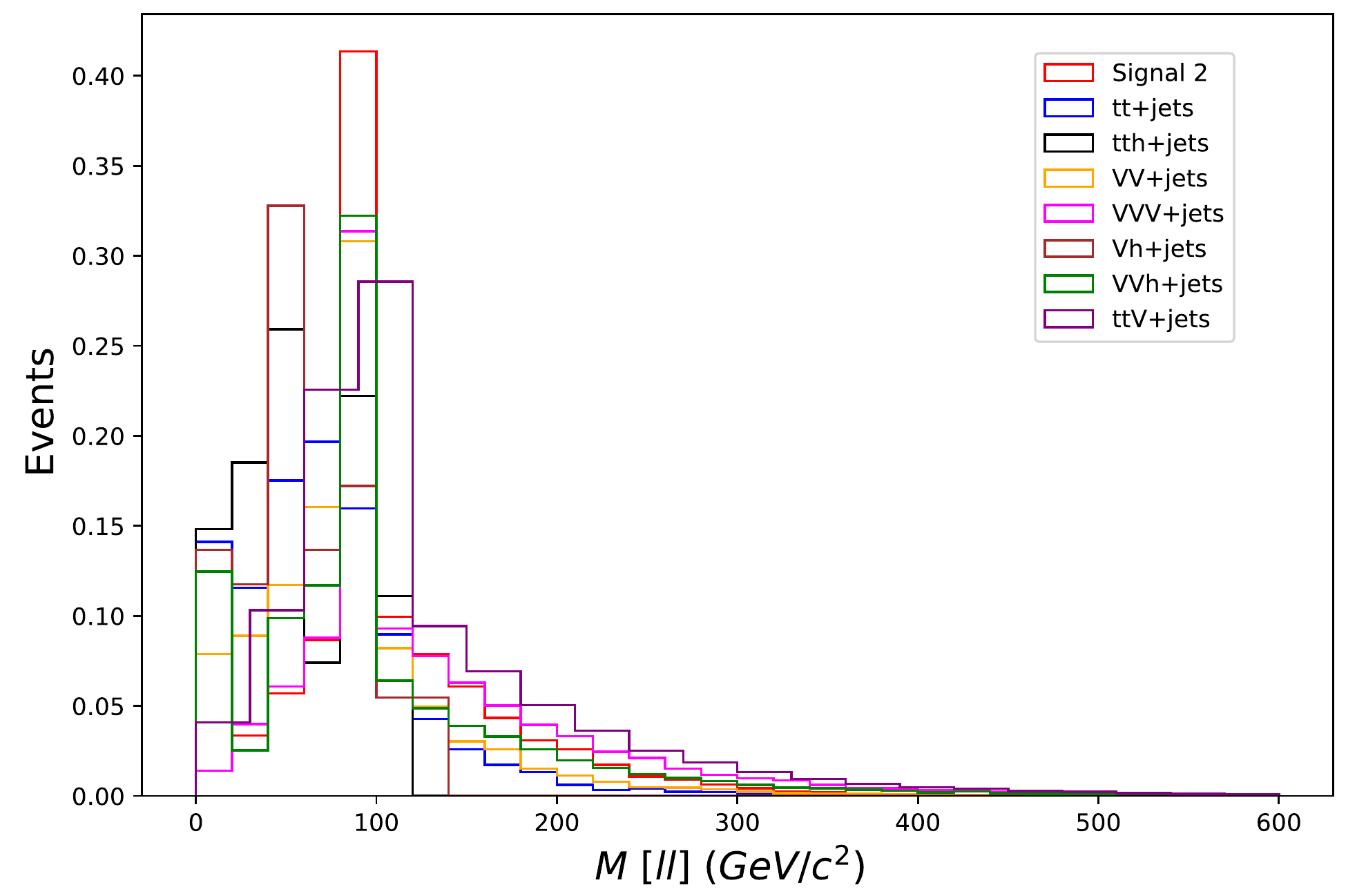}
\caption{The ${H}_T$ and $M_{\ell\ell}$ distribution for both signal and SM backgrounds with $m_{H^{\pm}}$ = 500 GeV and $m_{W^{'}}$ = 350 GeV for Signal 2.}
\label{fig:S2_HT}
\end{figure}

\begin{figure}[h!]
\includegraphics[scale=0.35]{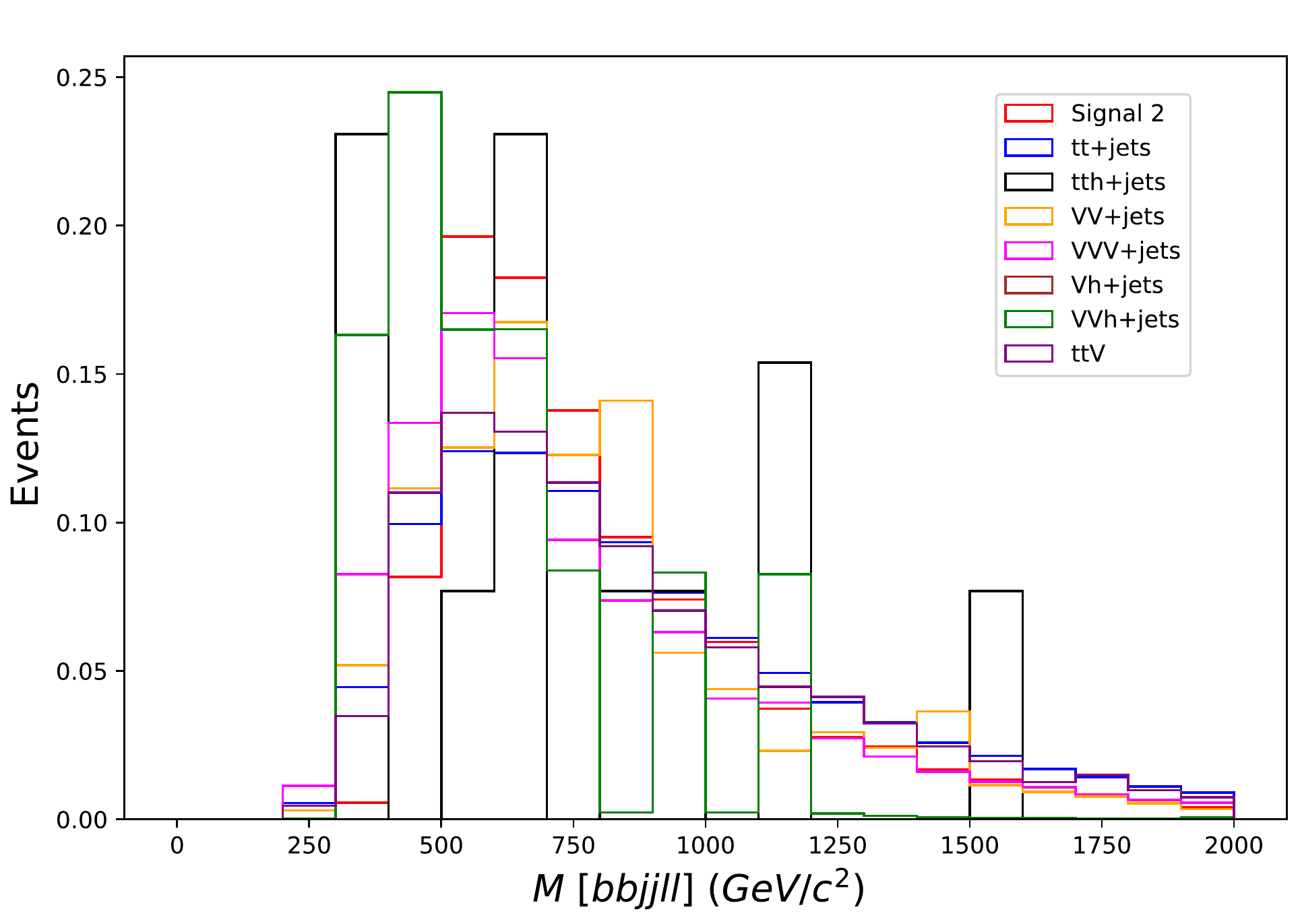}
\hspace{0.01in}
\includegraphics[scale=0.35]{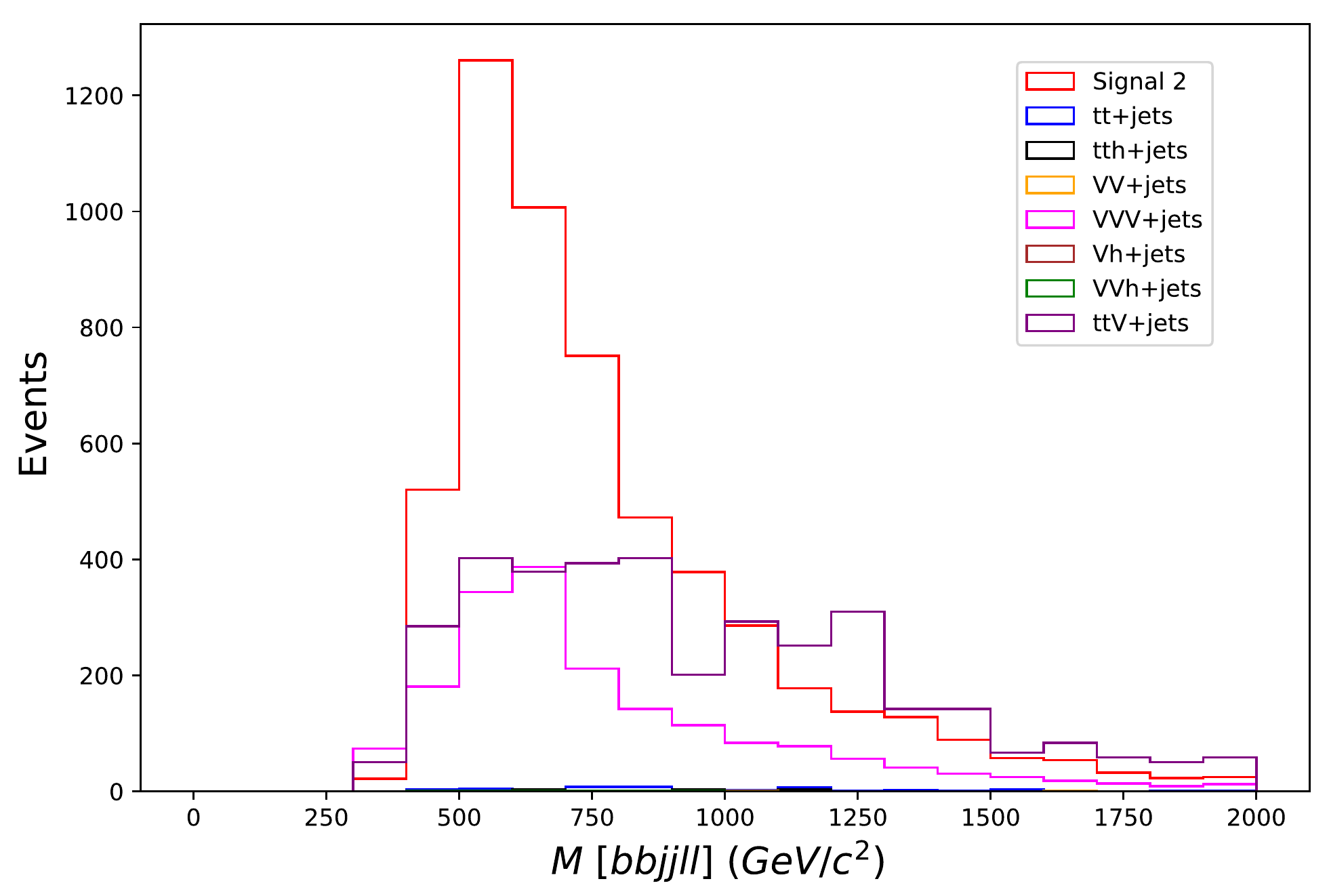}
\caption{The charged Higgs invariant mass distribution before and after implementing the cut mentioned in Table~\ref{tab:Sig2BP1}.}
\label{fig:S2_invariant}
\end{figure}

Since the $W'$ is assumed to have non-zero couplings to the scalar sector, we now consider the decay chain $pp \rightarrow H^{\pm} \bar{t} \rightarrow W^{'\pm} Z t  \,(W^{'\pm} \rightarrow W^{\pm} h) \rightarrow W^{\pm} W^{\mp} Z h \bar{b} \rightarrow 2j + 3\ell + 3b + \cancel{E}_{T} $ - we once again begin with  \textbf{BP1} The cut flowchart for this particular benchmark point for this signal is displayed in Table~\ref{tab:Sig2BP1}. We begin with a $H_T$ cut as before, and further demand at least 2 $b$-jets and 3 leptons. Further, we choose an invariant mass window of the two hardest leptons around the $Z$ mass. Since in this case both the signal and the $t\bar{t}$ background will have a sizeable missing energy, we have not put in a $\cancel{E}_T$ cut. The $H_T$ and $M_{\ell\ell}$ distributions for this case are shown in Fig.~\ref{fig:S2_HT} - it can be seen that the hadronic transverse momentum has the same qualitative features as that of Signal 1 thus explaining the same choice of cuts. The invariant mass distribution $m_{2b2j2\ell}$ before and after implementation of the cuts is shown in Fig.~\ref{fig:S2_invariant}. It can be seen that the choice of cuts has rendered most background negligible except for $VVV+\textrm{jets}$ - however this background has a rather small cross-section and should not provide a great impediment to discovery.

\begin{table}[h!]
	\centering
	\begin{tabular}{|c|c|c|c|c|c|c|c|c|}
		\hline
		Cuts & Signal            & $t\bar{t}$+jets            & $VV$+jets             & $t\bar{t}h$+jets &$VVV+$jets   & $Vh$+jets & $VVh$+jets & $t\bar{t}V$\\ \hline
		-      &  100000       & 4350000         &  2300000   &  100000 & 500000 & 190000 & 270000 & 200000 \\ \hline 
		$H_{T} \geq$ 400 GeV & 98145  & 1912953 & 426732 & 72508 &  285408 & 17125 & 176719 & 130377\\ \hline 
		$N_{b} \geq$ 2 & 70643  & 811669 & 153534 & 57603 & 173530 & 6557 & 139914 & 54744 \\ \hline
		$N_{\ell} \geq$ 3 & 9816  & 223 & 87 & 40 & 4659 & 1  & 27 & 7003\\ \hline
		70 GeV $\leq M_{\ell\bar{\ell}} \leq$ 120 GeV & 5511 & 50 & 26 & 13 & 1885 & 0 & 7 & 3812\\ \hline
	\end{tabular}
	\caption{ Cut flow chart for the $2j + 3l + 3b + \cancel{E}_T $ channel corresponding to $m_{H^{\pm}}$=500 GeV and $m_{W^{'}}$=350 GeV for Signal 2.}
	\label{tab:Sig2BP1}
\end{table}

In Table~\ref{tab:Sig2BP2}, we consider the \textbf{BP2} with a wider mass splitting as before. Comparing with Table~\ref{tab:Sig1BP2}, we have put in a $p_T$ cut on the leading jet in lieu of the missing energy cut. In addition, we have resorted to a tight(er) $m_{\ell\ell}$ cut to minimize backgrounds with vector bosons.

\begin{table}[h!]
	\centering
	\begin{tabular}{|c|c|c|c|c|c|c|c|c|}
		\hline
		Cuts & Signal             & $t\bar{t}$+jets            & $VV$+jets             & $t\bar{t}h$+jets &$VVV+$jets   & $Vh$+jets & $VVh$+jets & $t\bar{t}V$ \\ \hline
		-      &  100000        & 4350000         &  2300000   &  100000 & 500000 & 190000 & 270000 & 200000\\ \hline 
		$H_{T} \geq$ 600 GeV & 93955  & 668744 & 117293 & 35957 &  106306 & 4880 & 51502 & 57907\\ \hline 
		$N_{j} \geq$ 2  & 91061 & 654530 & 109727 & 33554 &  97829 & 4111 &  47473 & 55770 \\ \hline 
		$N_{\ell} \geq$ 3 & 22849  & 314 & 1633 & 19 & 5323 & 1 & 20 & 9060\\ \hline
		$N_{b} \geq$ 2 & 15382 & 88 & 28 & 16 & 533 & 0 & 10 & 3627  \\ \hline
		$p_{T}(\ell)  \geq$ 100 & 14385   &42 & 22 & 6 & 466 & 0 & 3 & 2948 \\ \hline
		80 GeV $\leq M_{\ell\bar{\ell}} \leq$ 100 GeV & 5378  & 7 & 7 & 6 & 125  & 0 & 1 & 1113\\ \hline
	\end{tabular}
	\caption{Cut-flow chart for the signal $2j + 3\ell + 3b + \cancel{E}_{T}$ channel with the signal corresponding to $m_{H^{\pm}} = 700$ GeV and $m_{W^{'}} = $350 GeV.}
	\label{tab:Sig2BP2}
\end{table}

\subsubsection{Signal 3}

We now move to the third process $p p \rightarrow H^{\pm} \bar{t} \rightarrow W^{'\pm} Z t (W^{'\pm} \rightarrow W^{\pm} Z) \rightarrow W^{\pm} W^{\mp} Z Z \bar{b} \rightarrow 4j + 2\ell + 3b$. Note that this is quite similar to Signal 1 and the difference arises from the way the SM gauge bosons decay. Here we require the $Z$ that comes from the $H^\pm$ to decay to $b\bar{b}$ while the other $Z$ (from the $W'$ decay) decays leptonically. While it is certainly advantageous to exploit the $b$-jet tagging at the LHC in addition to higher branching ratios, the $t\bar{t}+$jets background would need to be suppressed carefully. However, as we demonstrate in Table~\ref{tab:Sig3BP1} for  \textbf{BP1}, this can indeed be achieved. We find, as before, that the background events that remain after the $H_T$ and basic identification cuts can be efficiently reduced with a $\cancel{E}_T$ cut. In Table~\ref{tab:Sig3BP2}, we display the efficacy of cuts for the other benchmark point $m_{H^{\pm}} = 700$ GeV and $m_{W^{'}} = $350 GeV - it is seen that the final number of background events is smaller in this case owing to the harder $H_T$ cuts involved. Thus, the discovery of heavy or light charged Higgses in these channels is a delicate balance between getting enough signal events (which is difficult for heavier $H^\pm$) and suppressing the background more effectively (which is easier for heavier $H^\pm$). We revisit this issue in the next section.

\begin{table}[h!]
	\centering
	\begin{tabular}{|c|c|c|c|c|c|c|c|c|}
		\hline
		Cuts & Signal            & $t\bar{t}$+jets            & $VV$+jets             & $t\bar{t}h$+jets &$VVV+$jets   & $Vh$+jets & $VVh$+jets & $t\bar{t}V$ \\ \hline
		-      &  100000        & 4350000         &  2300000   &  100000 & 500000 & 190000 & 270000 & 200000 \\ \hline 
		$H_{T} \geq$ 400 GeV & 98956  & 1912953 & 426732 & 72508 &  285408 & 9013 & 176719 & 130377\\ \hline 
		$N_{j} \geq$ 4 & 88725  &1261763 & 197852 & 37199 & 143580 & 3234 & 86785& 76641 \\ \hline
		$N_{\ell} \geq$ 2 & 21210  & 360640 & 58065 & 8106 & 6812 & 357  & 115 & 35223 \\ \hline
		$N_{b} \geq$ 3 & 5905  & 11006 & 854 & 2007 & 30 & 5 & 6 & 1116\\ \hline
		$\cancel{E}_T \leq$ 50 GeV& 4403  & 2758 & 232 & 399 & 20 & 4  & 6 & 377\\ \hline
		80 GeV $\leq M_{\ell\bar{\ell}} \leq$ 100 GeV & 4116  & 339 & 43 & 83 & 3 & 4 & 0 & 162\\ \hline
	\end{tabular}
	\caption{Cut flow chart for the $4j + 2\ell + 3b $ channel with the signal corresponding to $ m_{H^{\pm}}$=500 GeV and $m_{W^{'}}$=350 GeV. }
	\label{tab:Sig3BP1}
\end{table}

\begin{table}[h!]
	\centering
	\begin{tabular}{|c|c|c|c|c|c|c|c|c|}
		\hline
		Cuts & Signal            & $t\bar{t}$+jets            & $VV$+jets             & $t\bar{t}h$+jets &$VVV+$jets   & $Vh$+jets & $VVh$+jets & $t\bar{t}V$\\ \hline
		-      &  100000       & 4350000         &  2300000   &  100000 & 500000 & 190000 & 270000 & 200000\\ \hline 
		$H_{T} \geq$ 600 GeV & 95024  & 668744 & 117293 & 35957 & 106306 & 2568 & 51502 & 57907\\ \hline 
		$N_{j} \geq$ 4 & 84330  & 514346 & 68189 & 22548 & 57965 & 1192 & 24477 & 38144 \\ \hline
		$N_{b} \geq$ 3 & 27083  & 33615 & 2560 & 8622 & 8382 & 52 & 6810 & 2009\\ \hline
		$MET \leq$ 50 & 17918  & 6884 & 549 & 1608 & 2864 & 26 & 2389 & 567\\ \hline
		$P_{T}(\ell) \geq$ 100 & 10977  & 1742 & 177 & 352 & 561 & 5 & 442 & 266 \\ \hline
		80 GeV $\leq M_{\ell\bar{\ell}} \leq$ 100 GeV & 5766  & 48 & 15 & 16 & 2 & 1 & 0 & 91 \\ \hline
	\end{tabular}
	\caption{Cut-flow chart for the signal $4j + 2\ell + 3b$ channel with the signal corresponding to $m_{H^{\pm}} = 700$ GeV and $m_{W^{'}} = $ 350 GeV.}
	\label{tab:Sig3BP2}
\end{table}

Finally before closing the section, we calculate the number of signal events necessary for discovery or exclusion for each signal events for both the benchmark points using \cite{Cowan:2010js}

\begin{align}
\begin{split}
&\mathcal{Z}_{D} = \sqrt{2\left[(S + B)\log\left[1 + \frac{S}{B}\right] - S\right]}, \\
&\mathcal{Z}_{E} = \sqrt{-2\left(B\log\bigg[1 + \frac{S}{B}\bigg] - S\right)},
\end{split}
\end{align}
where $S = \sigma_{S}\mathcal{L}$ and $B = \sigma_{B}\mathcal{L}$ are the total number of signal and background events that survive the cuts.  In Table~\ref{tab:cross-section}, we present the corresponding numbers for both the benchmark points and for all three signal scenarios discussed above.  

\begin{table}[h!]
\centering
\begin{tabular}{|c|c|c||c|c|c||c|c|c||c|c|c|}
\hline
\multirow{2}{*}{Signal} & \multirow{2}{*}{\textbf{BP}} &  \multirow{2}{*}{Background} & \multicolumn{3}{c||}{$\mathcal{Z}_{E} \geq$ 1.96} & \multicolumn{3}{c||}{$\mathcal{Z}_{D} \geq$ 3$\sigma$} & \multicolumn{3}{c|}{$\mathcal{Z}_{D} \geq$ 5$\sigma$}    \\ \cline{4-12} 
                      				&					&            & $\mathcal{L}$ = 500  & $\mathcal{L}$ = 1000 & $\mathcal{L}$ = 3000    & $\mathcal{L}$ = 500 & $\mathcal{L}$ = 1000 & $\mathcal{L}$ = 3000  & $\mathcal{L}$ = 500 & $\mathcal{L}$ = 1000 & $\mathcal{L}$ = 3000  \\ 
         &             &    in $fb$    &       $ fb^{-1} $ &  $ fb^{-1} $ & $ fb^{-1} $ & $ fb^{-1} $ & $fb^{-1}$ & $fb^{-1}$ & $ fb^{-1} $ & $fb^{-1}$ & $fb^{-1}$ \\ \hline	
         &  &  &  &  &  &  &  &  &  &  &	 \\							
               & \textbf{BP1} & 0.051 & 0.024 &0.155  & 0.09 & 0.034 & 0.024 & 0.013 & 0.058 & 0.041 & 0.022 \\
Signal 1 &       &          &           &         &     &  &  & & & &   \\
               		& \textbf{BP2} & 0.117 & 0.034 & 0.0244  & 0.0133 & 0.05 & 0.035 & 0.019 & 0.086 & 0.608 & 0.032\\
		\hline
				& & & & & & & & & & & \\
               & \textbf{BP1} & 1.619 & 0.12 & 0.085   &  0.046 & 0.174 & 0.123 & 0.071 & 0.294 & 0.207 & 0.118 \\
Signal 2 &       &          &           &         &     & & & & & &   \\
               		& \textbf{BP2} & 0.253 & 0.05 & 0.0353  & 0.02 & 0.072 & 0.051 & 0.028 & 0.122 & 0.0862 & 0.048 \\
		\hline
				& & & & & & & & & & & \\
               & \textbf{BP1} & 7.40 & 0.25 & 0.176 & 0.098 & 0.37 & 0.261 & 0.153 & 0.62 & 0.438 & 0.25\\
Signal 3 &       &          &           &         &   & & & & & &    \\
               		& \textbf{BP2} & 1.175 & 0.1 & 0.071 & 0.04 & 0.15 & 0.106 & 0.06 & 0.26 & 0.183 & 0.101 \\
		\hline													
  \end{tabular}
\caption{Estimated signal cross section for discovery and exculsion after imposing the cuts for both the benchmark points in all three signal scenarios.}
\label{tab:cross-section}
\end{table}
It can be seen that Signal  ($ 4j + 2\ell + 3b$) requires slightly lower values of signal cross-section for both exclusion (the $\mathcal{Z}_{E} \geq$ 1.96 column) and a $5\sigma$ discovery compared to the other two. However, the feasibility of one or the other depends on strength of couplings and patterns of decays and can only be answered within a model-specific context. We next turn to the issue of answering this in a toy model.

\section{Model Implications}
\label{sec:model}
\subsection{A Toy Model}

\begin{figure}[h!]
\begin{center}
\includegraphics[scale=0.36]{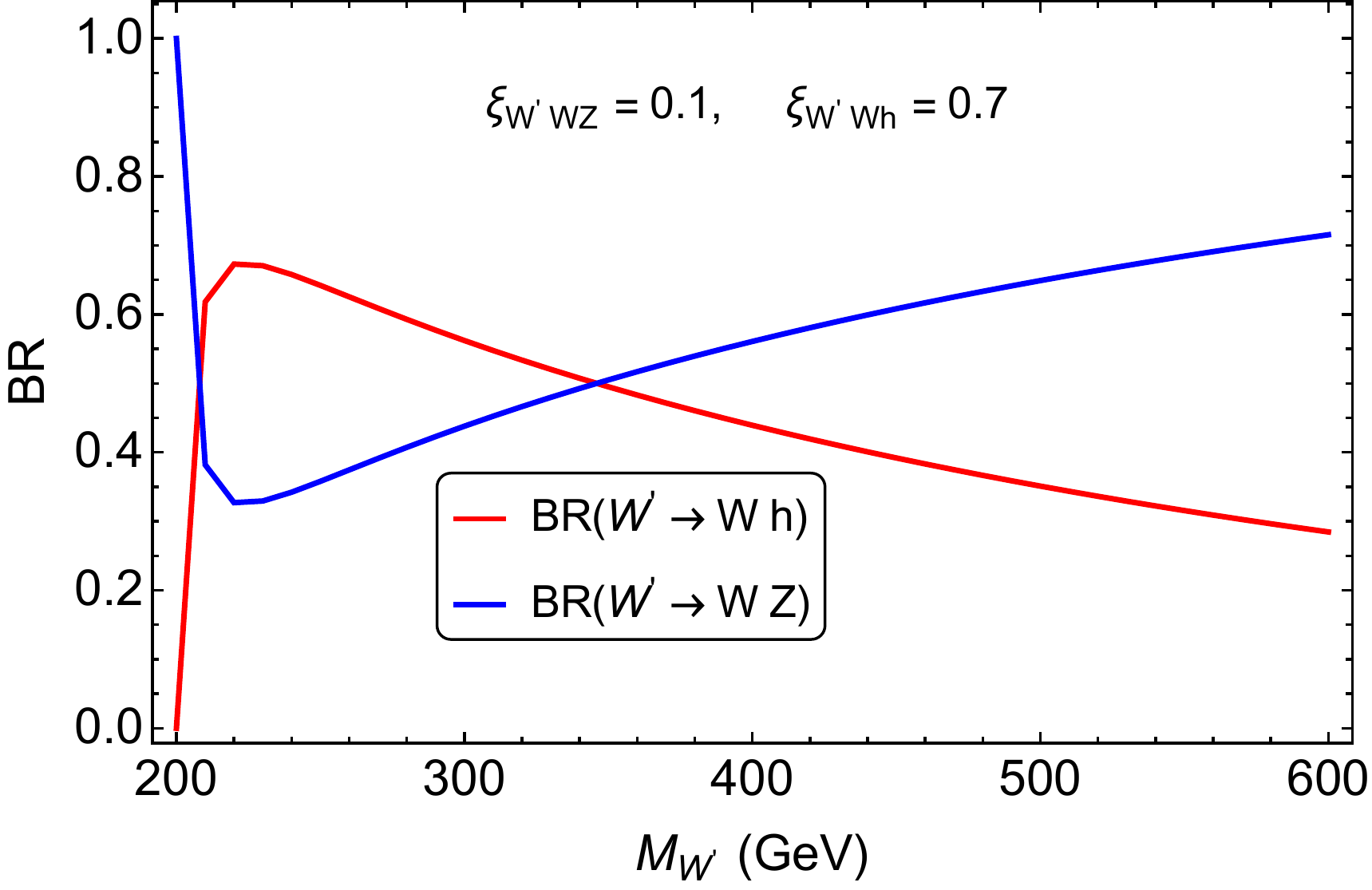}
\hspace{0.5in}
\includegraphics[scale=0.37]{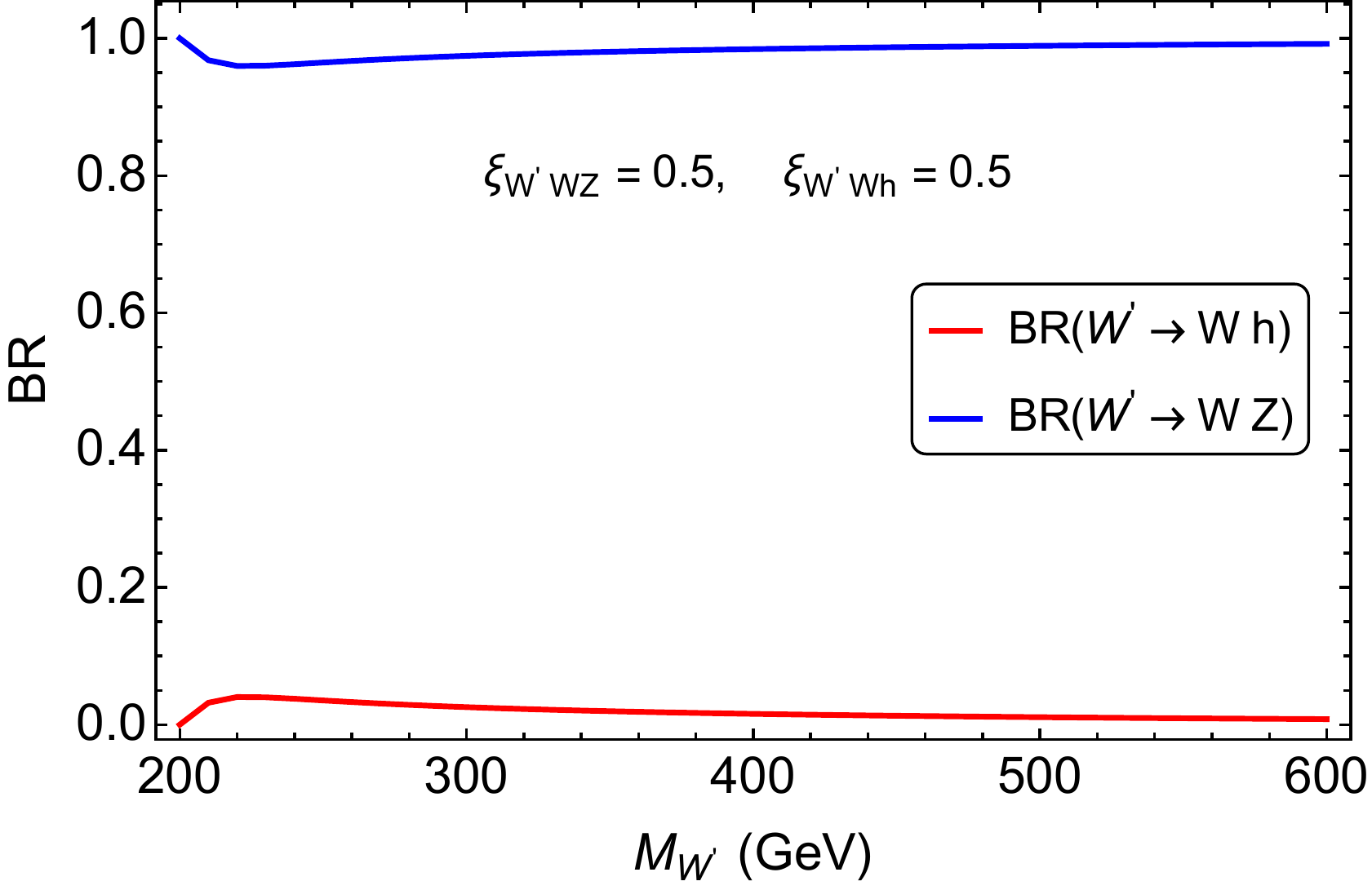}
\end{center}
\vspace{-0.75cm}
\caption{$W^{'}$ boson decay branching ratio for different choices of coupling strengths.}
\label{fig:Wprime}
\end{figure}

We begin our discussion with a toy model to get a sense of the numbers derived in Section~\ref{Sec:Pheno}. While a typical BSM scenario could have multiple particles and involved patterns of symmetry breaking, herein we concentrate on the minimal set necessary to illustrate our results along the lines of \cite{Cen:2018okf}. This includes the presence of a charged Higgs and a heavy $W'$ in addition to the SM particle spectrum. We begin with the relevant terms in the Lagrangian  for the $W'$.

\begin{align}
\mathcal{L}^{\textrm{int}}_{W^{'}} & = ~ \xi^{W'}_{WZ}\left[\partial_{\mu}W'_{\nu}\left(W_{\mu}Z_{\nu} - W_{\nu}Z_{\mu}\right) + W'_{\mu}\left( - W_{\nu}\partial_{\mu}Z_{\nu} + Z_{\nu}\partial_{\mu}W_{\nu} + W_{\nu}\partial_{\nu}Z_{\mu} - Z_{\nu}\partial_{\nu}W_{\mu} \right) \right] \nonumber \\
	& ~~ + \xi^{W'}_{VS}\left[m_{W'}W'_{\mu}W_{\mu}S \right] + \textrm{h.c.},
\label{Eq:WpLag}
\end{align}
where $S$ denotes a generic neutral scalar and $m_{W^{'}}$ is the mass of the additional gauge boson. The interaction terms between the SM and the new gauge bosons come about after rewriting the gauge kinetic energy in terms of the mass eigenstates in the usual manner. For our current purposes, we restrict to the case where the masses of the possible additional scalars (other than the SM-like Higgs boson) are heavier than the $W^{'}_{\mu}$. In addition, since we are dealing with a fermiophobic $W'$\footnote{It is certainly possible to relax this criterion and consider a $W'$ with sufficiently weak couplings to the SM fermions to evade the direct search limits and remain light.}, we do not consider decays to SM quarks and leptons. Thus the only relevant decay channels are the $WZ$ and the $Wh$ modes - we write down the corresponding decay widths below:    

\begin{center}
\begin{equation}
\Gamma[W' \rightarrow W Z] = \left( \frac{\xi^{2}_{W'WZ}m^{5}_{W'}}{192\pi m^{2}_{W}m^{2}_{Z}}\right)\left[ 1 + 10\left( \frac{m^{2}_{12}}{m^{2}_{W'}} \right) + \left(\frac{m_{12}^{4} + 8m^{2}_{W}m^{2}_{Z}}{m^{4}_{W'}}\right) \right]\sqrt{\left( 1 - \frac{m_{+}^{2}}{m^{2}_{W'}} \right)^{3}\left( 1 - \frac{m_{-}^{2}}{m^{2}_{W'}}\right)^{3}},
\label{Eq:WpWZ}
\end{equation}

\begin{equation}
\Gamma[W' \rightarrow W h] = \left(\frac{\xi^{2}_{W'Wh}m^{2}_{W}m_{W'}}{12\pi v^{2}} \right)\left[ 2 + \frac{(m^{2}_{W'} + m^{2}_{W} - m^{2}_{h})^{2}}{4m^{2}_{W}m^{2}_{W'}}\right]\sqrt{1 - 2\frac{m^{2}_{W} + m^{2}_{h}}{m^{2}_{W'}} + \frac{(m^{2}_{h} - m^{2}_{W})^{2}}{m^{4}_{W'}}},
\label{Eq:WpWh}
\end{equation}
\end{center}
where $m_{12} = m^{2}_{W} + m^{2}_{Z}$ and  $m_{\pm} = m_{W} \pm m_{Z}$.  The couplings $\xi_{W'WZ}$ are $\xi_{W'Wh}$ are taken as free parameters (these typically depend on the new gauge couplings in an extended gauge model), and we fix them to reasonable $\mathcal{O}(1)$ numbers in what follows. In Fig~\ref{fig:Wprime}, we display the branching ratio of the $W'$ in the mass range $200-600$ GeV for two different choices of the couplings. One can notice that even for low values of $\xi_{W'WZ}$, the $WZ$ channel dominates over the $Wh$ for $m_{W'}>350$ GeV.   

\begin{figure}[h!]
\begin{center}
\includegraphics[scale=0.4]{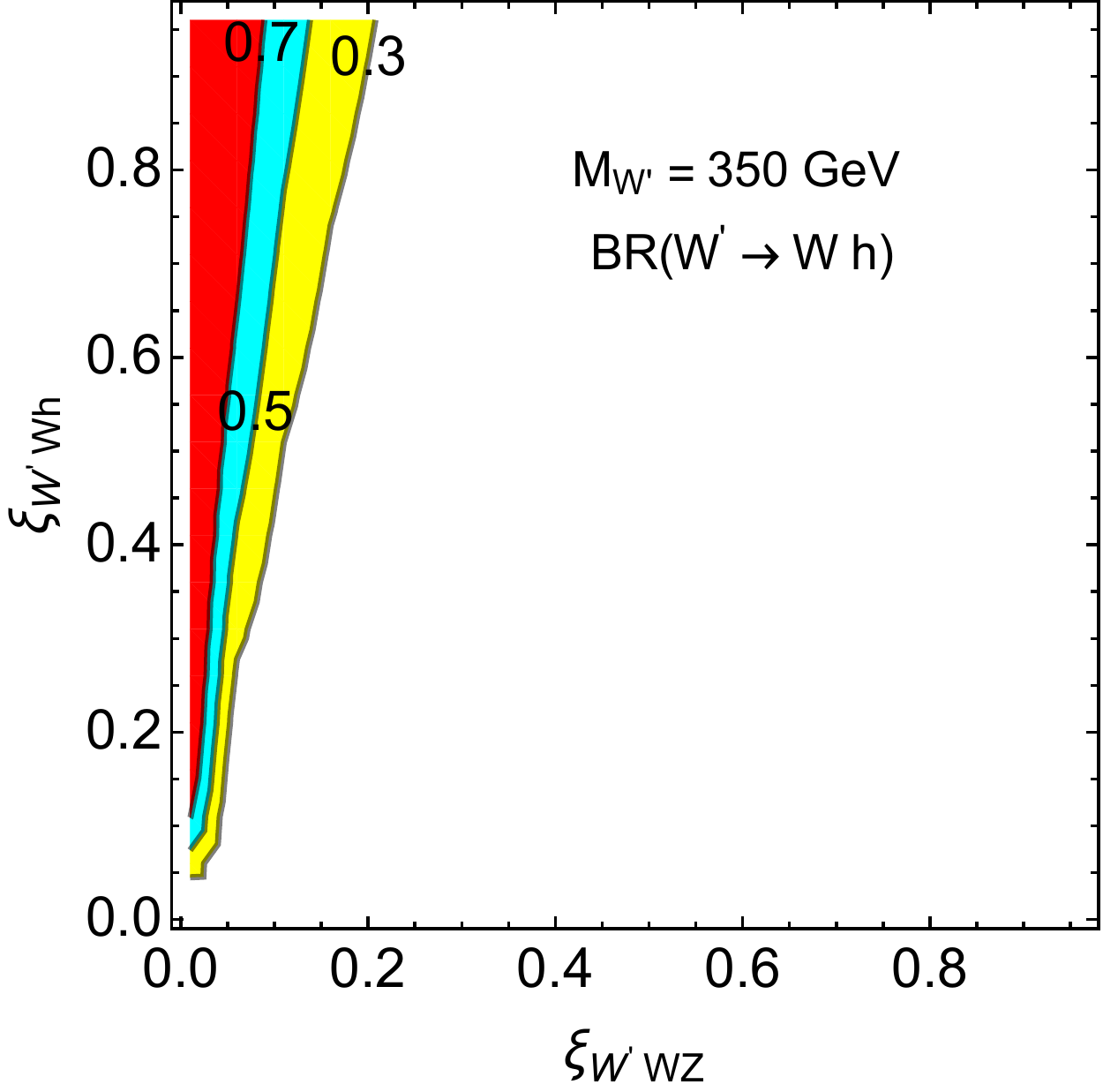}
\hspace{0.5in}
\includegraphics[scale=0.4]{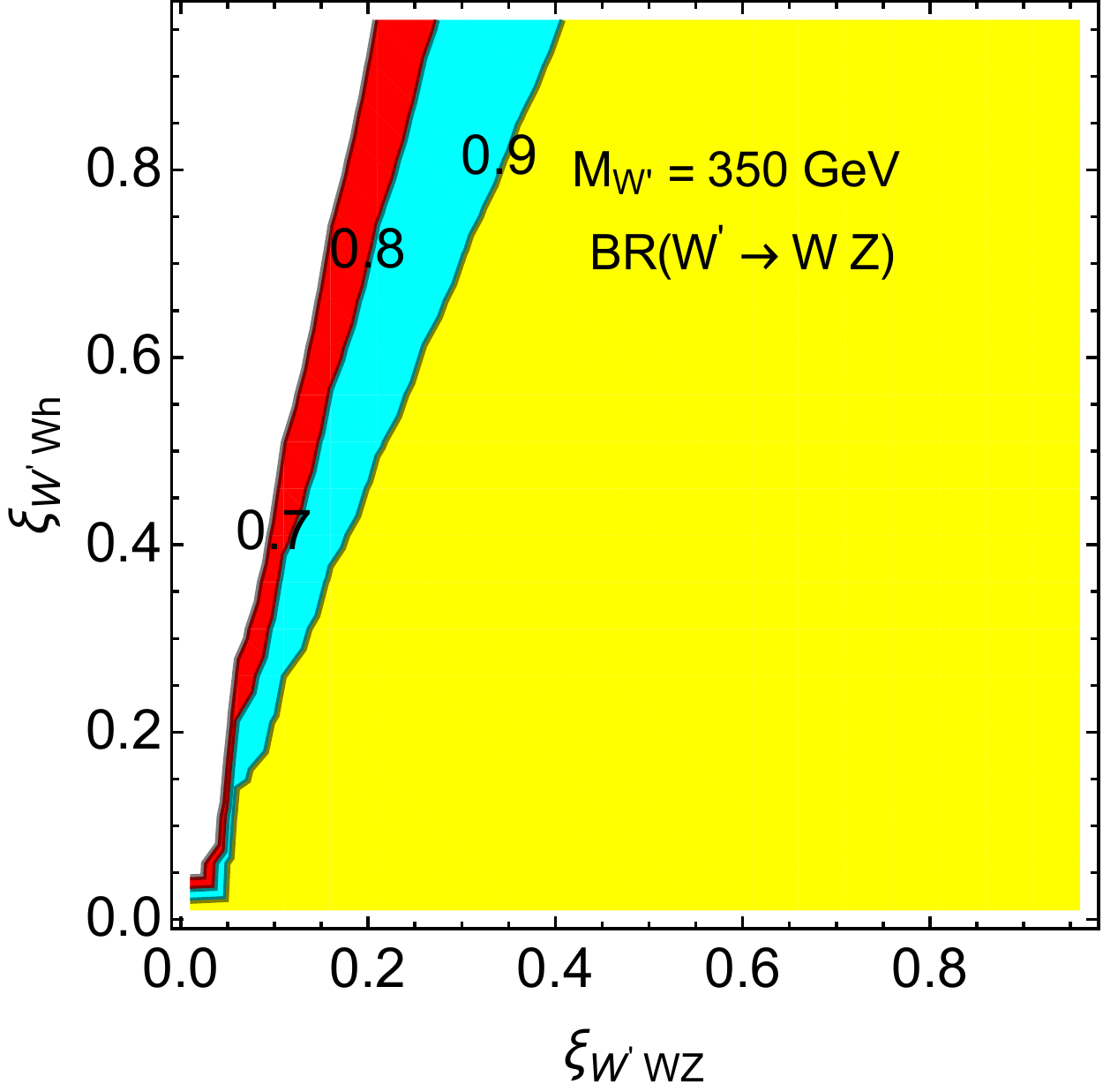}
\end{center}
\caption{Contours of branching ratio for the $Wh$ (left) and the $WZ$ (right) channels for a heavy $W'$ of mass 350 GeV.}
\label{fig:WpCont}
\end{figure}

In Fig~\ref{fig:WpCont}, we illustrate the variation in branching ratio values for both these decay modes for $m_{W'}=$ 350 GeV. For $\xi_{W'WZ} \geq$ 0.4, $\textrm{BR}( W' \rightarrow WZ)$ is greater than 90\% for entire range of $\xi_{W'Wh}$. On the other hand, the $\textrm{BR}(W' \rightarrow Wh)$ has an appreciable value (say, $\gtrsim$ 30\%) in the region $\xi_{W'WZ} <$  0.2. In our subsequent analysis, we will fix $\textrm{BR}(W' \rightarrow W Z)$ to 0.95  (for Signals 1 and 3) and                    $\textrm{BR}(W' \rightarrow W h)$ to 0.7 (for Signal 2) as these numbers represent the best case scenarios under the present considerations. 

Turning now to the charged Higgs boson, in Eqn~\ref{Eq:Hceffect}, we write down the most general phenomenological Lagrangian that describes the interaction of the $H^\pm$ with the SM sector.

\begin{equation}
\mathcal{L}^{\textrm{int}}_{H^{\pm}} = \xi^{H^{\pm}}_{VS}H^{\pm}\left(V^{\mp}_\nu\partial_{\mu}S - S\partial_{\mu}V_{\nu}^{\mp}\right)g^{\mu\nu} + m_{V^{'}}\xi^{H^{\pm}}_{VV^{'}}H^{\pm}V^{'}_{\mu}V^{\mu} + H^{\pm}\bar{f}\left(\frac{m_{f}}{v}\xi^{H^{\pm}}_{f}P_{L} + \frac{m_{f^{'}}}{v}\xi^{H^{\pm}}_{f^{'}}P_{R}\right)f^{'} + \textrm{h.c.},
\label{Eq:Hceffect}
\end{equation}
where the projection operators are defined in the usual way: $P_{L/R} = \frac{(1 \mp \gamma_{5})}{2}$,  and $\xi^{H^{\pm}}_{VS}$, $\xi^{H^{\pm}}_{f/f^{'}}$, $\xi^{H^{\pm}}_{V^{'}V}$ are model dependent coupling parameters. In the first term, $V_\mu^{\pm}$ can denote either $W_\mu^{\pm}$ or its heavy counterpart (the couplings in the two cases would, of course, be different as is evident from the notation). The second term denotes interactions of the $H^\pm$ with pairs of gauge bosons $WZ$, $W'Z$, and $WZ'$ each with a coupling generically denoted by $\xi^{H^{\pm}}_{VV^{'}}$. While one can design models in which these couplings are independent, we assume the following pattern (with the assumption that the charged Higgs belongs in a multiplet that contributes to electroweak symmetry breaking): $\xi^{H^{\pm}}_{WZ^{'}}$ = $\cos^{2}\theta_{W}\xi^{H^{\pm}}_{W^{'}Z}$, $\xi^{H^{\pm}}_{WZ}$ = $\left( \frac{m_{W} }{ m_{W^{'}}}\right) \xi^{H^{\pm}}_{W^{'}Z}$, and similarly in the scalar sector $\xi^{H^{\pm}}_{W^{'}h}$ = $\left(\frac{m_{W}}{m_{W^{'}}}\right)\xi^{H^{\pm}}_{Wh}$. In general, the charged Higgs boson can couple to up- and down-type quarks and to leptons differently. However to reduce the number of free parameters and make our analysis simpler, we assume $\xi^{H^{\pm}}_{f}$ = $\xi^{H^{\pm}}_{f^{'}}$ = $\xi^{H^{\pm}}_{ff^{'}}$ (for example, like in the Type-I 2HDM \cite{Posch:2010hx}\cite{Branco:2011iw}).  Thus, we are left with three independent coupling parameters ($\xi^{H^{\pm}}_{W^{'}Z}$, $\xi^{H^{\pm}}_{ff^{'}}$ and $\xi^{H^{\pm}}_{Wh}$) that we will treat as free parameters in what follows and fix them at reasonable $\mathcal{O}(1)$ numbers. In Eqn~[\ref{Eq:HcVS} - \ref{Eq:HcVVp}], we present the decay widths for the charged Higgs in the various available channels.

\begin{equation}
\Gamma(H^{\pm} \rightarrow V S) = \left(\frac{m^{3}_{H^{\pm}}|\xi^{H^{\pm}}_{VS}|^{2}}{16\pi v^{2}}\right)\left[1 - \frac{(m_{V} + m_{S})^{2}}{m^{2}_{H^{\pm}}} \right]^{\frac{3}{2}}\left[1 - \frac{(m_{V} - m_{S})^{2}}{m^{2}_{H^{\pm}}} \right]^{\frac{3}{2}},
\label{Eq:HcVS}
\end{equation}

\begin{equation}
\Gamma(H^{\pm} \rightarrow f f^{'}) = \left(\frac{N_{c}\lambda^{\frac{1}{2}}(m^{2}_{H^{\pm}}, m^{2}_{f}, m^{2}_{f^{'}})}{8\pi v^{2}m^{3}_{H^{\pm}}}\right)\left[(m^{2}_{H^{\pm}} - m^{2}_{f^{'}} - m^{2}_{f})(m^{2}_{f^{'}} + m^{2}_{f})|\xi^{H^{\pm}}_{ff^{'}}|^{2} - 4m^{2}_{f}m^{2}_{f^{'}} \right], \textrm{and}
\label{Eq:Hcffp}
\end{equation}

\begin{equation}
\Gamma(H^{\pm} \rightarrow V V^{'}) = \left(\frac{m^{2}_{V}m^{2}_{V^{'}}|\xi^{H^{\pm}}_{VV^{'}}|^{2}}{4\pi v^{2}m_{H^{\pm}}}\right)\left(2 + \frac{(m^{2}_{H^{\pm}} - m^{2}_{V} - m^{2}_{V^{'}})}{4m^{2}_{V}m^{2}_{V^{'}}}\right)\sqrt{1 - 2\left(\frac{m^{2}_{V} + m^{2}_{V^{'}}}{m^{2}_{H^{\pm}}}\right) + \left(\frac{m^{2}_{V} - m^{2}_{V^{'}}}{m^{2}_{H^{\pm}}}\right)^{2}}.
\label{Eq:HcVVp}
\end{equation}

\begin{figure}[h!]
\begin{center}
\includegraphics[scale=0.31]{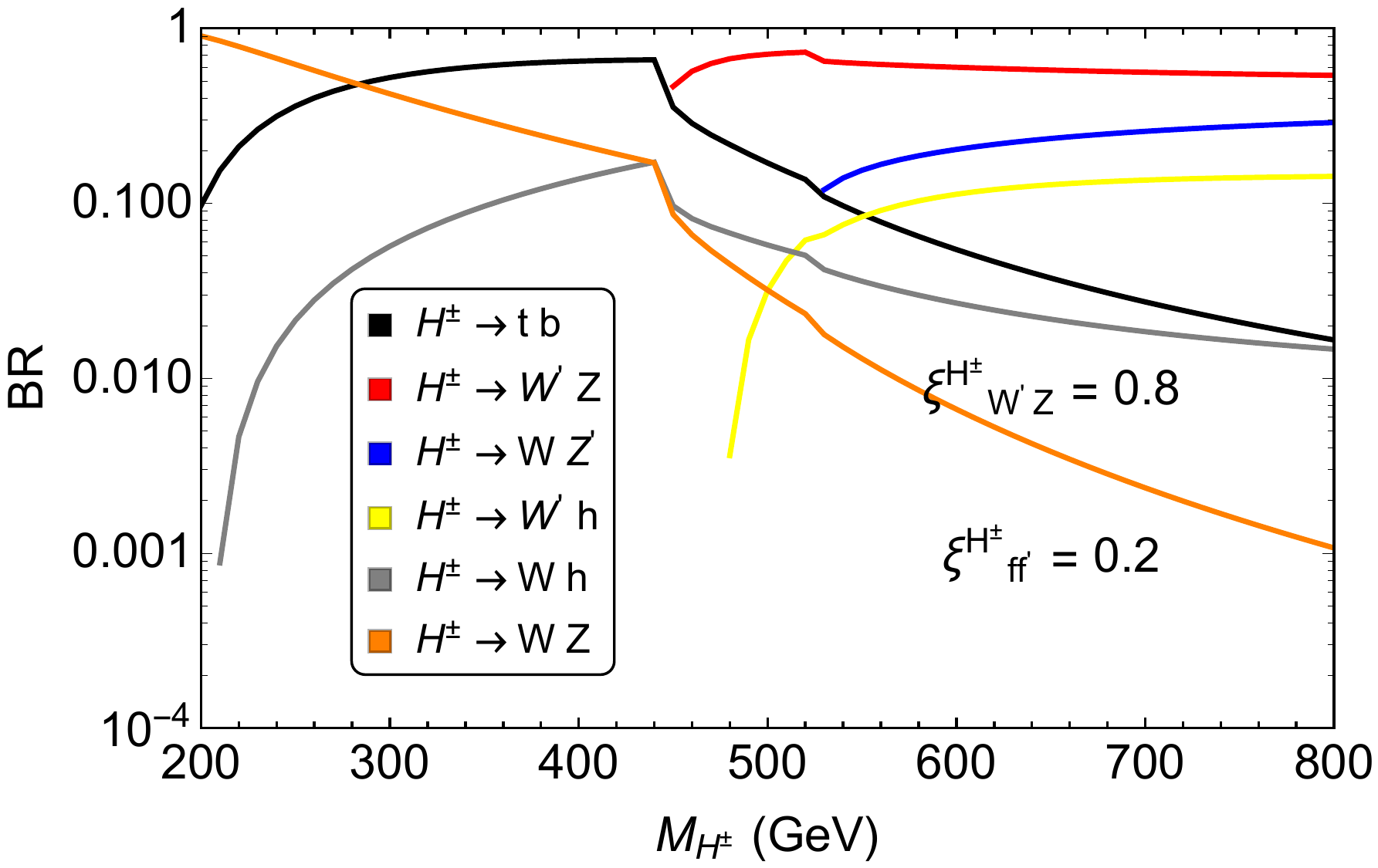}
\hspace{0.01in}
\includegraphics[scale=0.3]{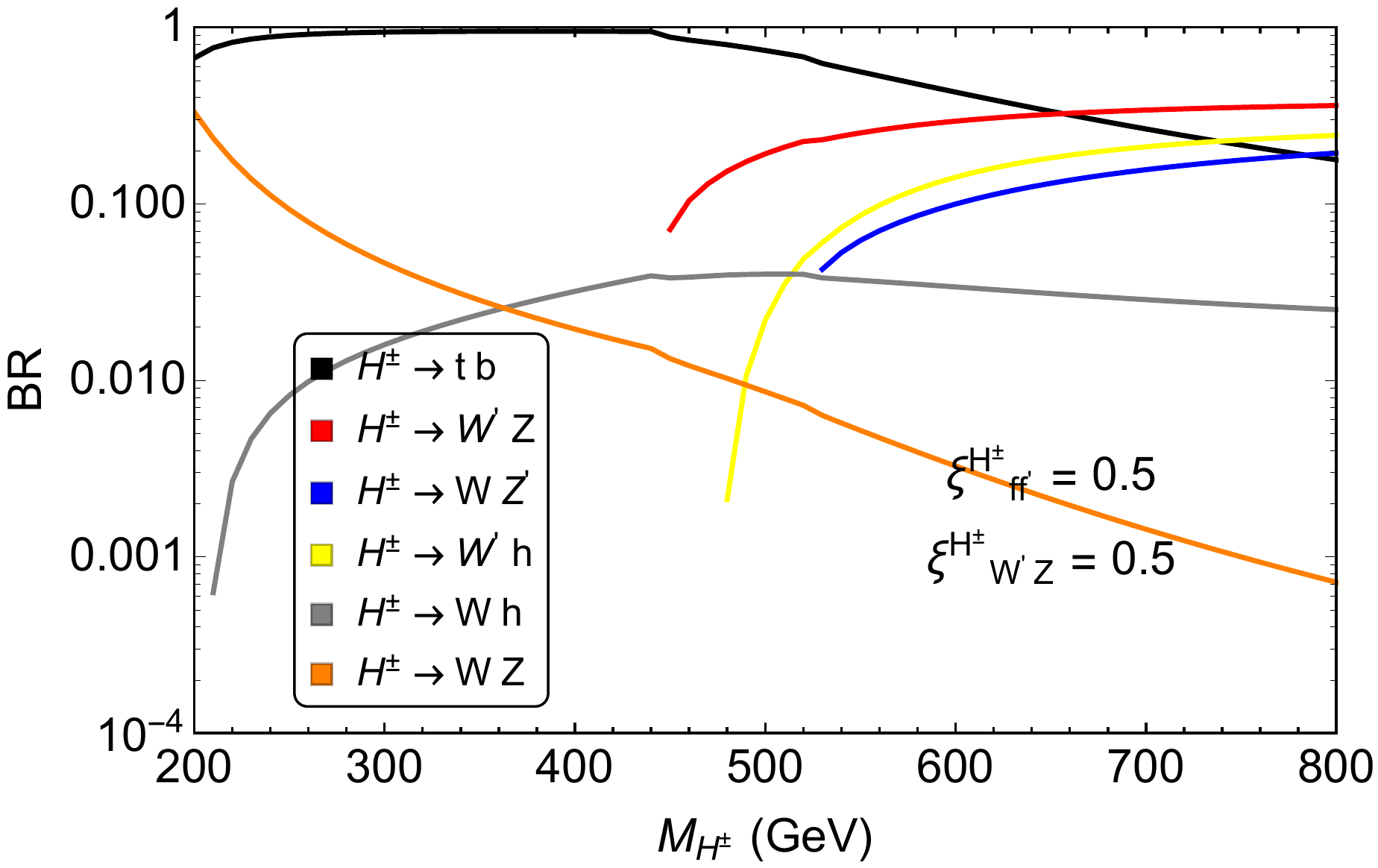}
\hspace{0.01in}
\includegraphics[scale=0.3]{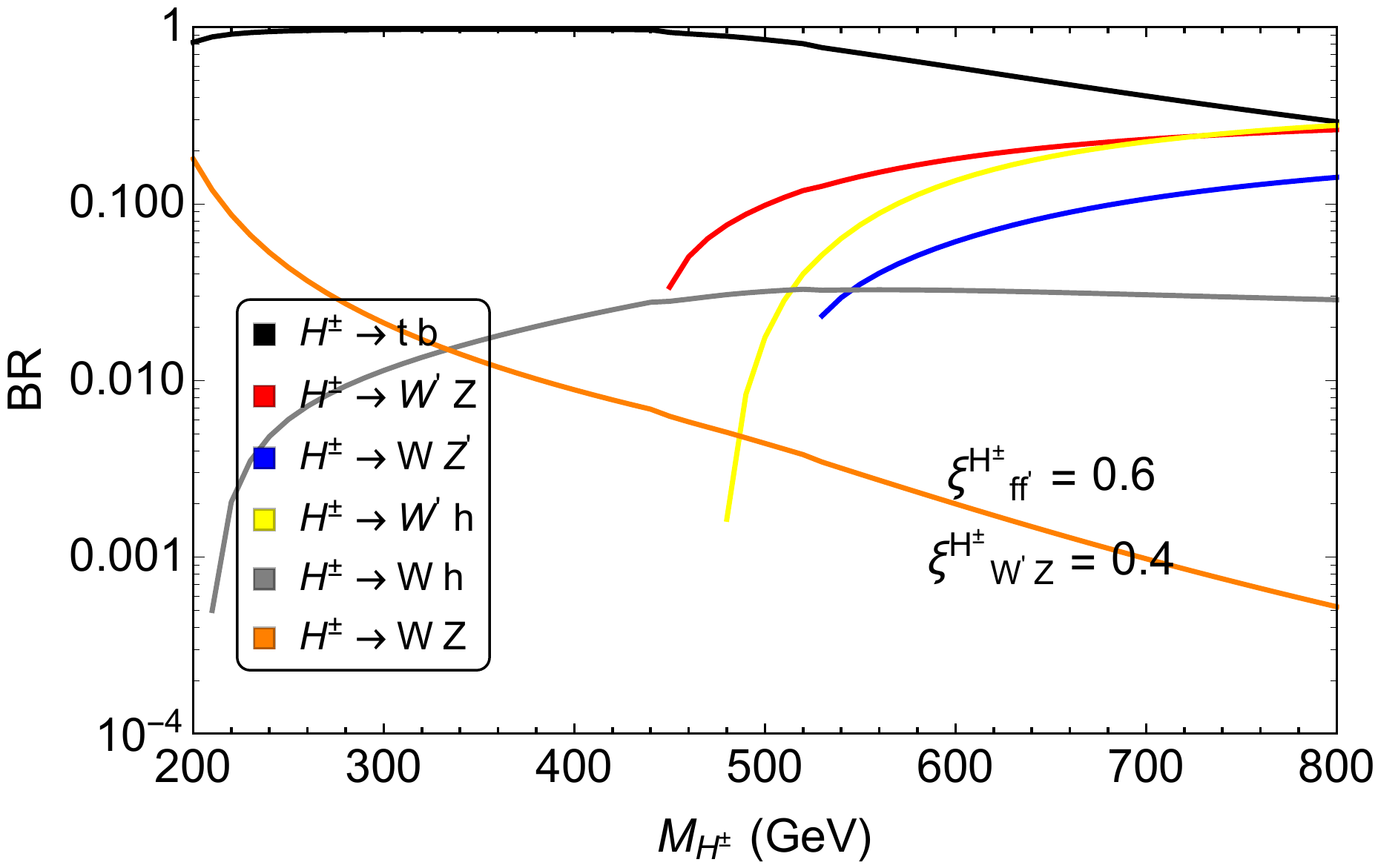}
\end{center}
\vspace{-0.5cm}
\caption{The charged Higgs boson branching ratio in the various allowed channels for different choices of coupling strength fixing $\xi^{H^{\pm}}_{Wh}$ = 0.1.}
\label{fig:HcDecay}
\end{figure}

In Fig~\ref{fig:HcDecay}, we show the $H^{\pm}$ boson decay branching ratios in different channels for various choices of $\xi^{H^{\pm}}_{W'Z}$ and $\xi^{\pm}_{ff^{'}}$ fixing $\xi^{\pm}_{Wh}$ at 0.1. One can observe that for moderate values of $\xi^{H^{\pm}}_{W^{'}Z}$, the $\textrm{BR}(H^{\pm} W^{'} Z)$ is non-negligible and can reach $\gtrsim$ 20\% for $m_{H^{\pm}} \geq$ 500 GeV. For slightly larger values of $\xi^{H^{\pm}}_{W'Z}$, this number can be as high as 50\% as is demonstrated in the first of the three plots in Fig.~\ref{fig:HcDecay}. Interestingly, the $H^{\pm} \rightarrow W Z'$ decay channel also has an appreciable contribution for $m_{H^{\pm}}\gtrsim$ 650 GeV. Moreover, if the $Z'$ decays to $Zh$, the cut flow chart presented for Signal 2 can be applicable.

\begin{figure}[h!]
\begin{center}
\includegraphics[scale=0.35]{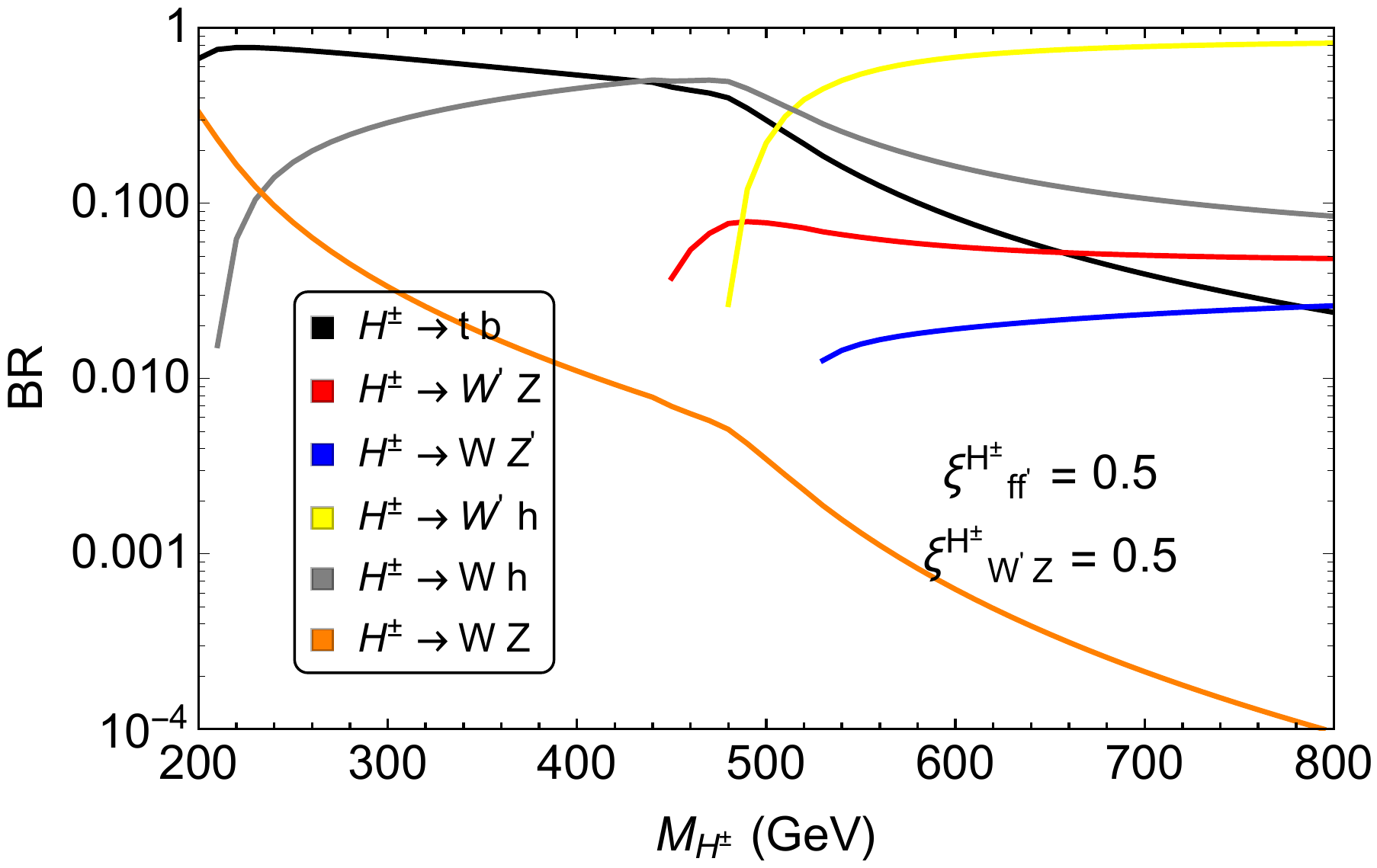}
\end{center}
\vspace{-0.5cm}
\caption{The charged Higgs boson branching ratio in the various allowed channels for $\xi^{H^{\pm}}_{ff^{'}} = \xi^{H^{\pm}}_{W^{'}Z} = \xi^{H^{\pm}}_{Wh}$ = 0.5.}
\label{fig:HcDecay_PartI}
\end{figure}

In Fig~\ref{fig:HcDecay_PartI}, we display the branching ratio for a democratic choice of couplings: $\xi^{H^{\pm}}_{ff^{'}} = \xi^{H^{\pm}}_{W^{'}Z} = \xi^{H^{\pm}}_{Wh}$ = 0.5. Interestingly in this case, for $m_{H^{\pm}} \geq$ 550 GeV, the $H^{\pm} \rightarrow W' h$ becomes the dominant decay mode. If we choose to decay the $W'$ to  $W Z$, then the final state topology in this case is identical to that of Signal 2. If we trade the $M_{\ell\bar{\ell}}$ cut with something like 110 GeV $\geq M_{b\bar{b}} < 140$ GeV cut in Table~\ref{tab:Sig2BP1}, a majority of the SM backgrounds can be rejected. However, the $t\bar{t}h$+jets background would be a challenge in this particular scenario and thus our cut flowchart for Signal 2 might have to be supplanted with additional cuts. If on the other hand we consider $W'\rightarrow W h$, the final state would be $WWhh + \bar{b}$. Though in the present study, we do not explore the collider phenomenology for these interesting channels and stick to the scenario where $\xi^{H^{\pm}}_{Wh}$ = 0.1, we mention them here to give an idea of the kind of rich phenomenology that the charged Higgs can enjoy in many extended gauge models.      

\begin{figure}[h!]
\begin{center}
\includegraphics[scale=0.4]{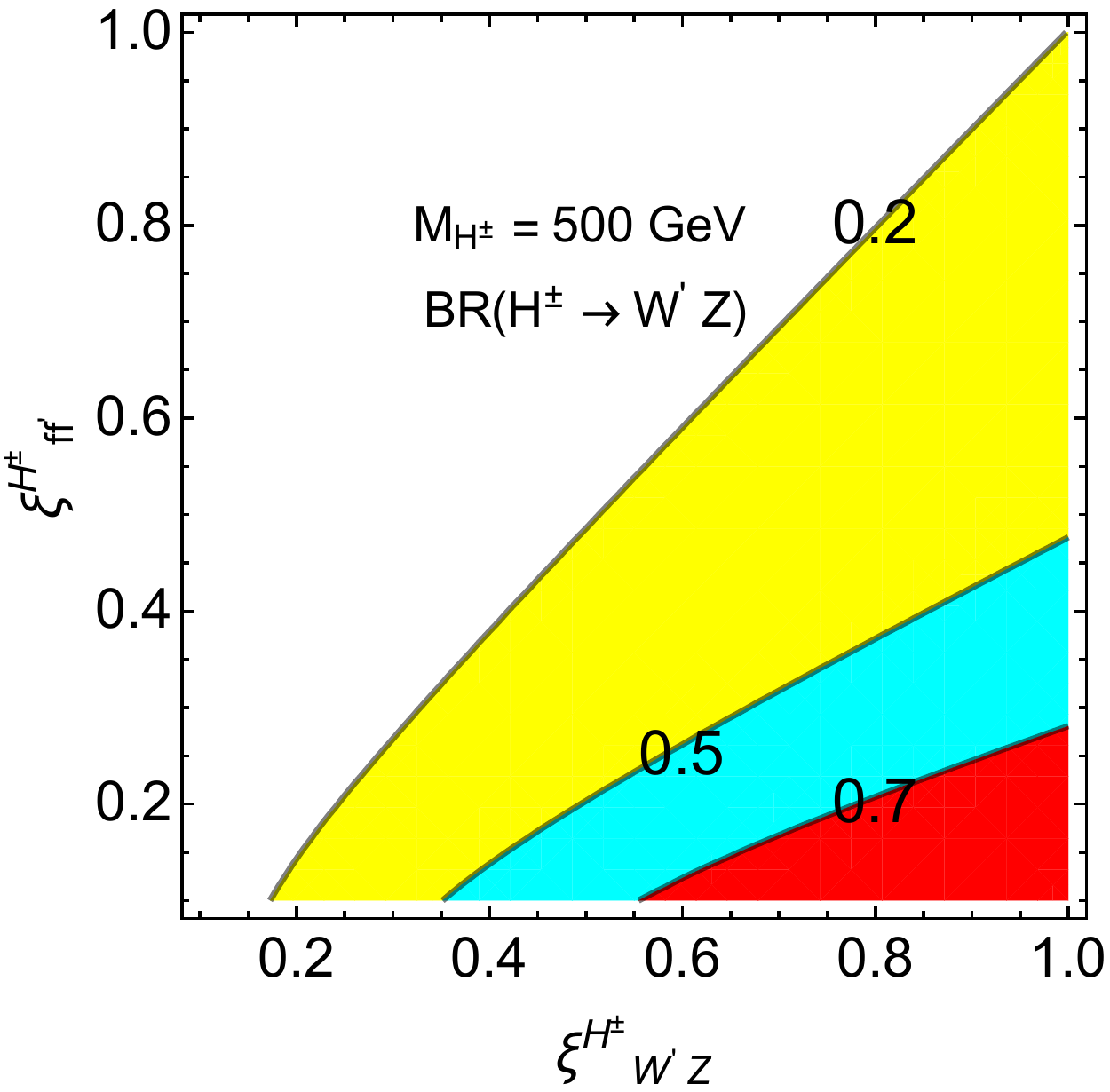}
\hspace{0.5in}
\includegraphics[scale=0.4]{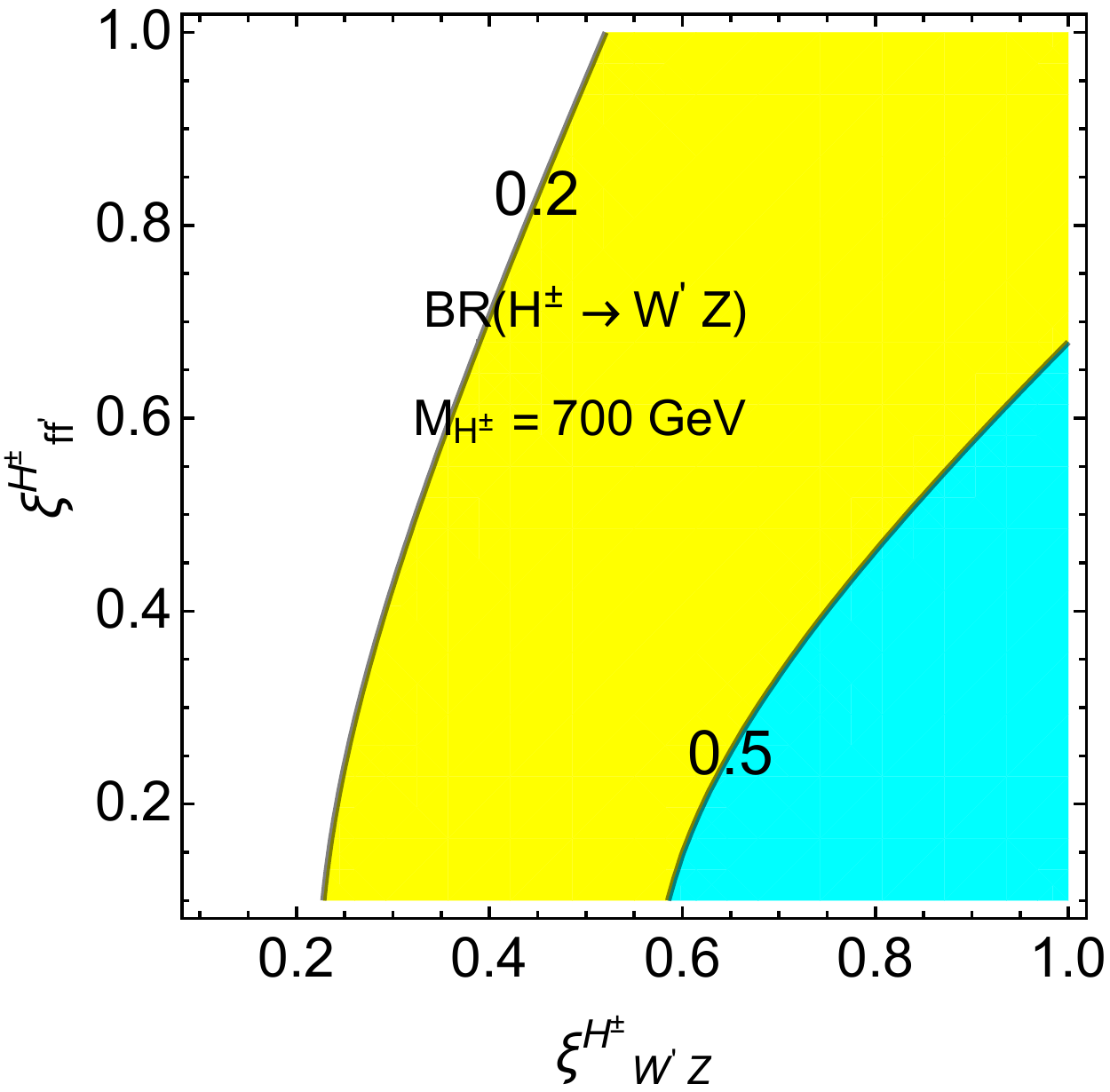}
\end{center}
\caption{The variation in the branching ratio for the mass value $m_{H^{\pm}}$ = 500 GeV and 700 GeV. The $\xi^{\pm}_{Wh}$ is fixed at 0.1.}
\label{fig:HcCont}
\end{figure}

To illustrate the differences between the two benchmark points, in Fig~[\ref{fig:HcCont}], we present the variation of $\mathcal{BR}(H^{\pm} \rightarrow W Z)$ in the parameter plane $\xi^{H^{\pm}}_{W^{'}Z}$ vs $\xi^{H^{\pm}}_{ff^{'}}$ plane for the two different choices of $m_{H^\pm}$ used in this study. A heavier $H^\pm$ decays to $W'h$ more readily and at the cost of the $W'Z$, and hence we find that this branching ratio cannot reach the numbers that the 500 GeV case can.

\begin{figure}[h!]
\begin{center}
\includegraphics[scale=0.35]{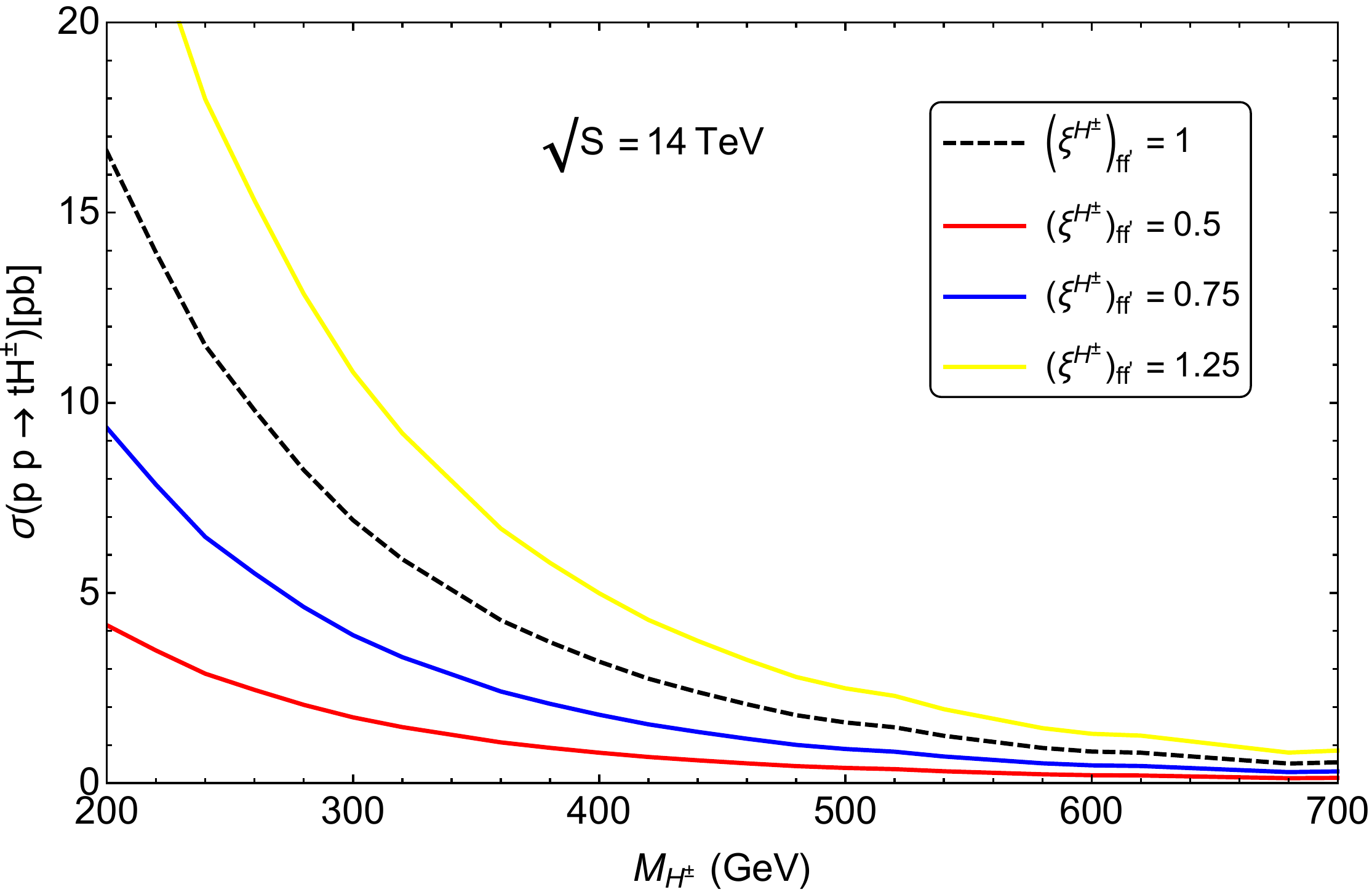}
\end{center}
\caption{The black dashed line in the plot represent the model independent production cross section for $p p \rightarrow H^{\pm}t$ for charged Higgs boson in between the  mass range 200 GeV to 600 GeV.}
\label{fig:gbHct}
\end{figure}

Any estimate finally has to involve both the production cross-section ($pp\to H^\pm \bar{t}$ in our case) and the various branching ratios. In Fig[\ref{fig:gbHct}] we present the production cross-section for the charged Higgs boson in the mass range  200 GeV to 700 GeV at the 14 TeV LHC via the associated production mode for various choices of $\xi^{H^{\pm}}_{ff^{'}}$. The black dashed line represents the cross-section values \cite{Plehn:2002vy,Chivukula:2011ag,Flechl:2014wfa} for $\xi^{H^{\pm}}_{ff^{'}}=1$.

\begin{figure}[h!]
\begin{center}
\includegraphics[scale=0.4]{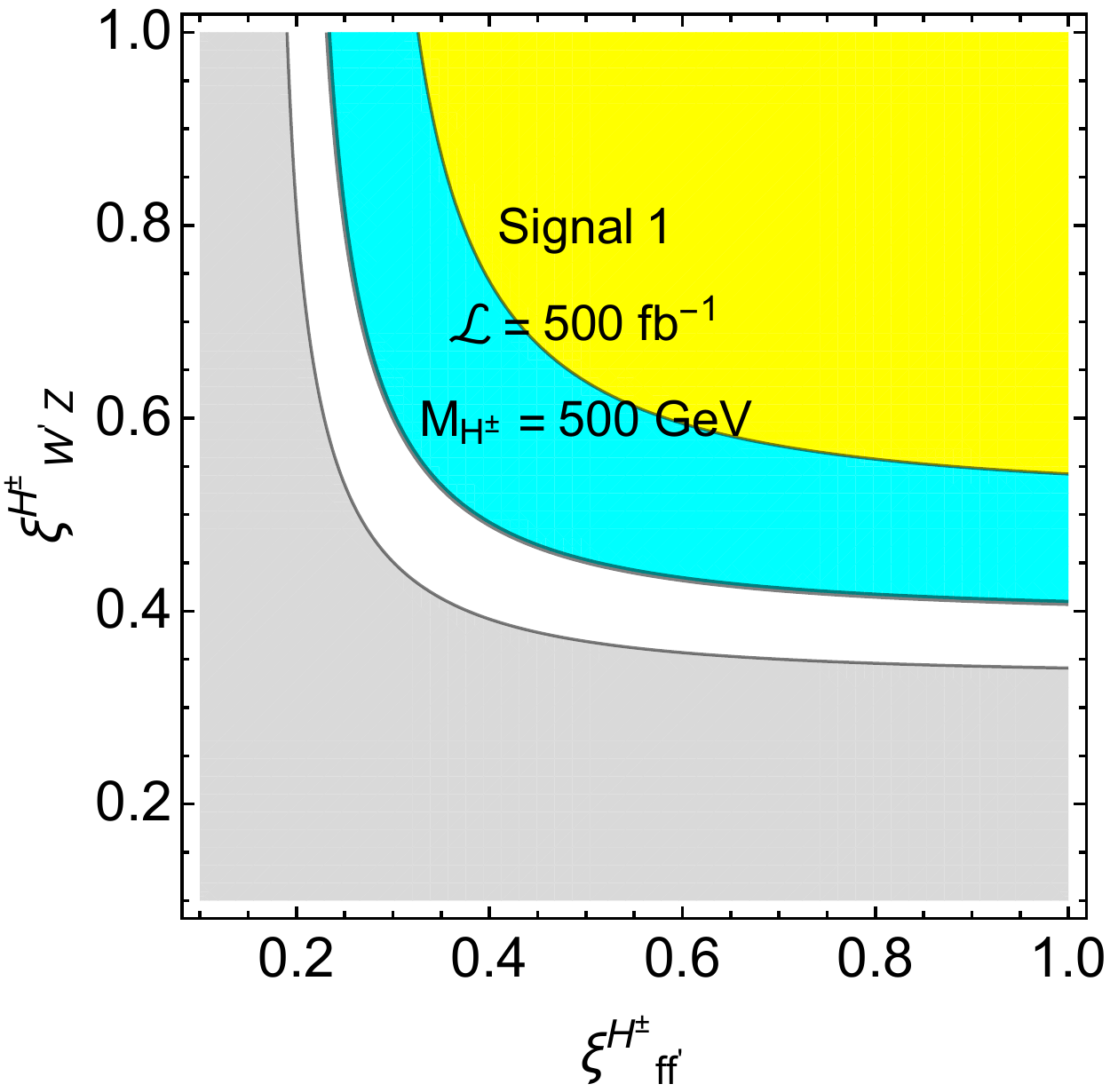}
\hspace{0.1in}
\includegraphics[scale=0.4]{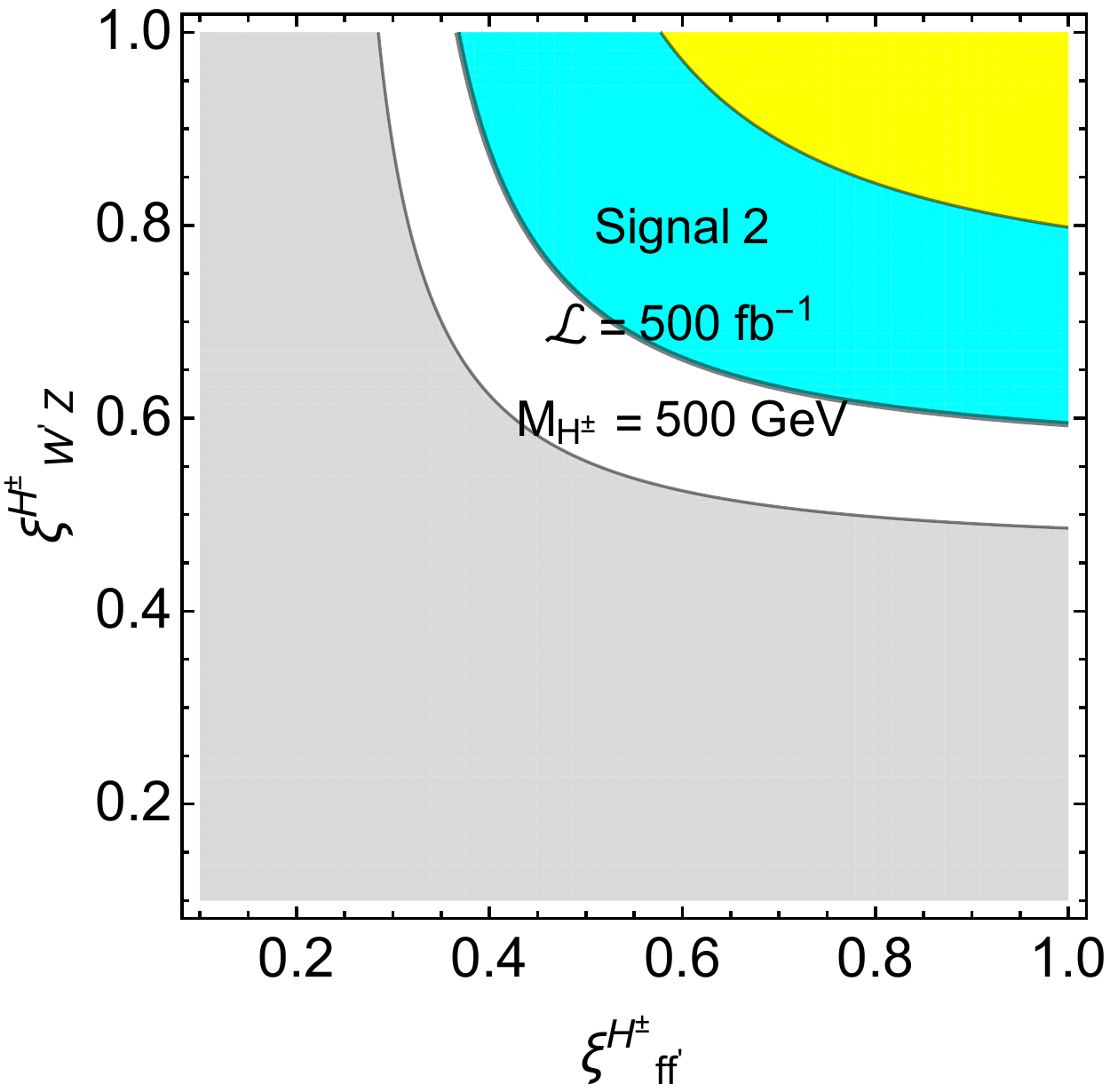}
\hspace{0.1in}
\includegraphics[scale=0.4]{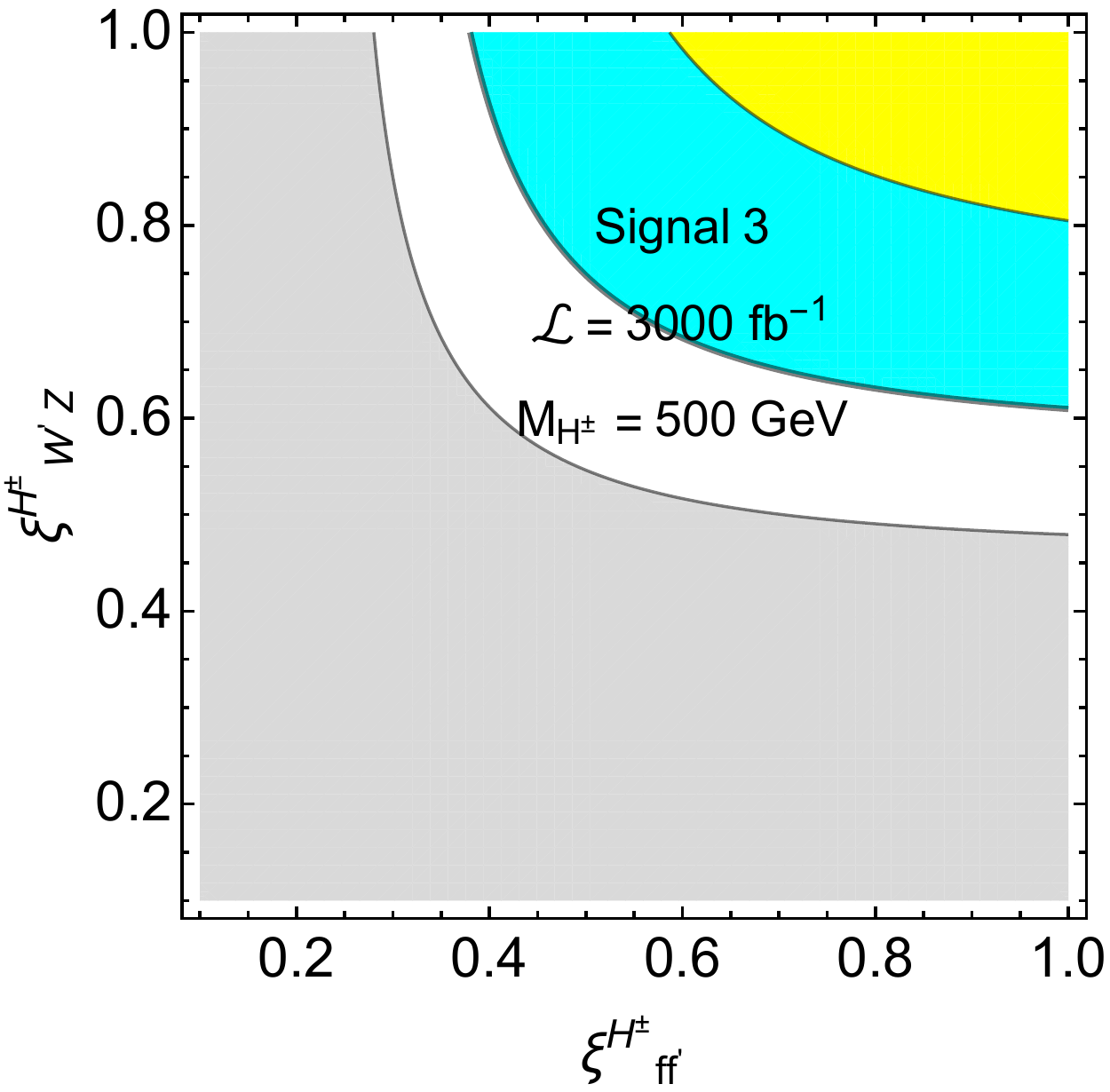}
\end{center}
\caption{The reach plot for different signal in the parameter plane $\xi^{H^{\pm}}_{W^{'}Z}$ versus $\xi^{H^{\pm}}_{tb}$ plane.}
\label{fig:HcCont_Param}
\end{figure}

Having specified the cross-section and the branching ratios, we are now in a position to present the reach plot for the $H^\pm$ in the various channels within the context of this toy model. For this purpose, we require
\begin{equation}
\epsilon_{s}\left[|\xi^{H^{\pm}}_{ff^{'}}|^{2}\sigma(g b \rightarrow H^{\pm} t)|_{\xi^{H^{\pm}}_{ff^{'}} = 1}\right]\mathcal{BR}(H^{\pm} \rightarrow W^{\pm}Z) \geq \mathcal{Z}_{D},
\end{equation}
where $\epsilon_{s}$ denotes efficiency of the cut based analysis, a number that is independent of the model dependent parameters. We fix $\epsilon_{s}$ to a rather modest value of 0.25 in the reach plots presented in Fig.~\ref{fig:HcCont_Param} - it should be noted that more parameter space will open up for slightly more liberal choices of efficiency. It can be seen that both Signal 1 and Signal 2 can prove to be quite effective in aiding the discovery of a charged Higgs, while Signal 3 requires higher luminosities. In the plot, the yellow and cyan regions correspond to $5\sigma$ and $3\sigma$ discovery regions while the gray region can potentially be excluded. While regions of $5\sigma$ discovery are admittedly small (particularly for Signal 3), it is clear from these plots that even in this toy model scenario, the extended gauge sector does offer new possibilities for a charged Higgs boson discovery that could be probed at the LHC.

\section{A Not So Toy Model}
\label{Model_TTM}

In a recent work \cite{Coleppa:2020set}, we proposed a BSM scenario with an enlarged gauge symmetry  $SU(2)_{0}\times SU(2)_{1}\times U(1)_{2}$. Symmetry breaking in this model is engineered by two Higgs doublets $\Phi_{1,2}$ and a non linear Sigma field $\Sigma$. The vev of $\Sigma$ and $\Phi_2$ is denoted by $F$ and that of $\Phi_1$ is denoted by $f$ and these are parametrized as 

\begin{equation}
F = \sqrt{2}v\cos\beta ~~~~~~~ f = v\sin\beta
\end{equation}   
The coupling constants of the $SU(2)_{0}\times SU(2)_{1}\times U(1)_{2}$ model are denoted by $g_{0}$, $g_{1}$ and $g_{2}$ respectively. After symmetry breaking the scalar spectrum of the model contains two CP-even Higgs boson ($H$, $h$), a pseudoscalar $A$, and a pair of charged Higgs bosons $H^{\pm}$. The Higgs-fermion couplings in this model mimics the traditional Type-I 2HDM at leading order. We identify the lighter CP-even Higgs as the $h-125$ GeV SM-like Higgs. The mixing angle between the two CP even Higgs is denoted by $\alpha$. The charged Higgs couplings in this model relevant for our discussion is presented in Table~\ref{tab:Hc_Coupling} in terms of the parameter $x=m_W/m_{W'}$.

\begin{table}[h!]
\centering
\begin{tabular}{|c||c|}
\hline
	&	\\
$\xi^{H^{\pm}}_{W^{'\mp}Z}$ & $\frac{\sin\beta}{2}\left(1 + \frac{x^{2}}{4}\right)$ \\
	&	\\
\hline
	& 	\\
$\xi^{H^{\pm}}_{W^{\mp}Z}$ & $\frac{x^{2}\cos\beta\sin\beta}{16\sin^{2}\theta_{w}\cos^{2}\theta_{w}}$ \\
	&	\\
\hline
	& 	\\
$\xi^{H^{\pm}}_{W^{\mp}Z^{'}}$ & $\frac{\sin\beta\cos\theta_{w}}{2}\left(1 + \frac{x^{2}}{8}\right)$ \\
	&	\\
\hline
	&	\\
$\xi^{H^{\pm}}_{ff^{'}}$ & $\cot\beta\left(1 - \frac{x^{2}}{4}\right)$ \\
	&	\\
\hline
	&	\\
$\xi^{H^{\pm}}_{W^{\mp}h}$ & $\frac{1}{4}\left[(4\sin\alpha\cos\beta + \sqrt{2}\sin\alpha\sin\beta) + \frac{x^{2}}{32}(8\sin\alpha\cos\beta - \sqrt{2}\sin\alpha\sin\beta) \right]$   \\	
	&	\\
\hline
	&	\\
$\xi^{H^{\pm}}_{W^{'\mp}h}$ &  $\frac{1}{\sqrt{2}x}\left[\sin\alpha\sin\beta - \frac{x^{2}}{32}(4\sin\alpha\cos\beta + \sqrt{2}\sin\alpha\sin\beta) \right]$ \\
\hline 			
\end{tabular}
\caption{The charged Higgs couplings of Ref~\cite{} that are relevant to the present phenomenological discussion.}
\label{tab:Hc_Coupling}
\end{table}

\begin{figure}[h!]
\begin{center}
\includegraphics[scale=0.45]{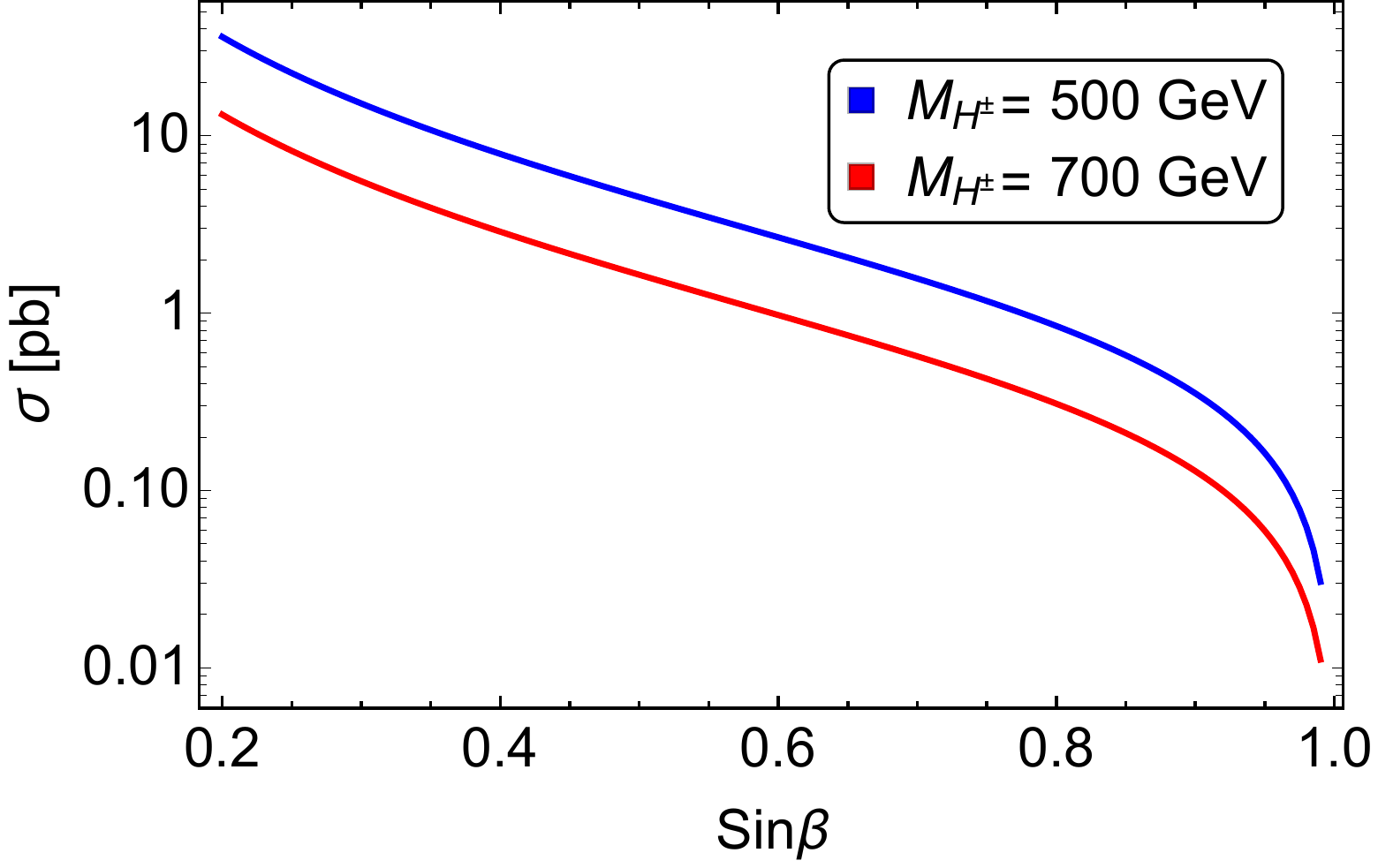}
\end{center}
\caption{The production cross-section for $p p \rightarrow H^{\pm}t$ for different values of $\sin\beta$.}
\label{fig:gbHct_Sinb}
\end{figure}

\begin{figure}[h!]
\begin{center}
\includegraphics[scale=0.348]{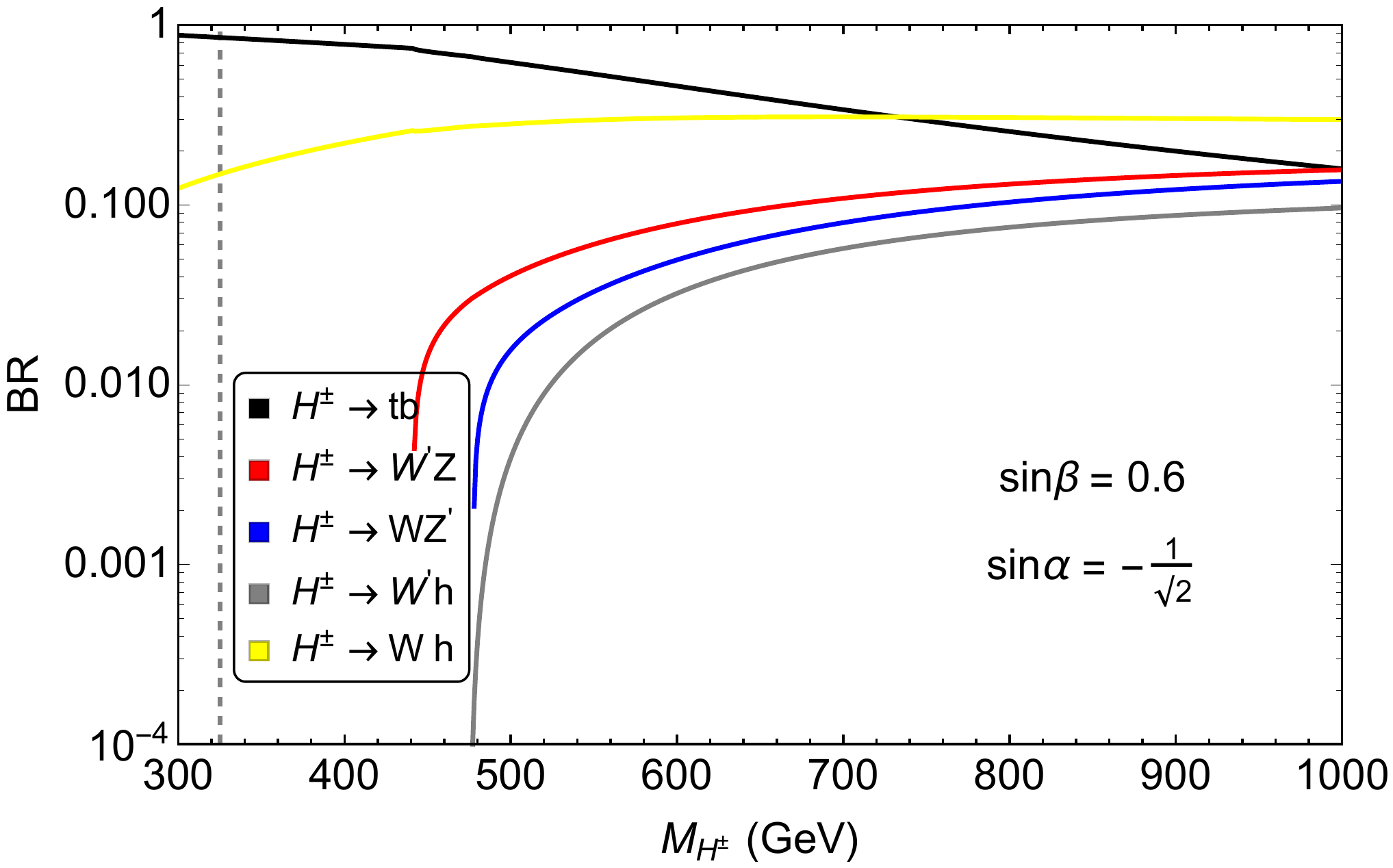}
\hspace{0.2in}
\includegraphics[scale=0.355]{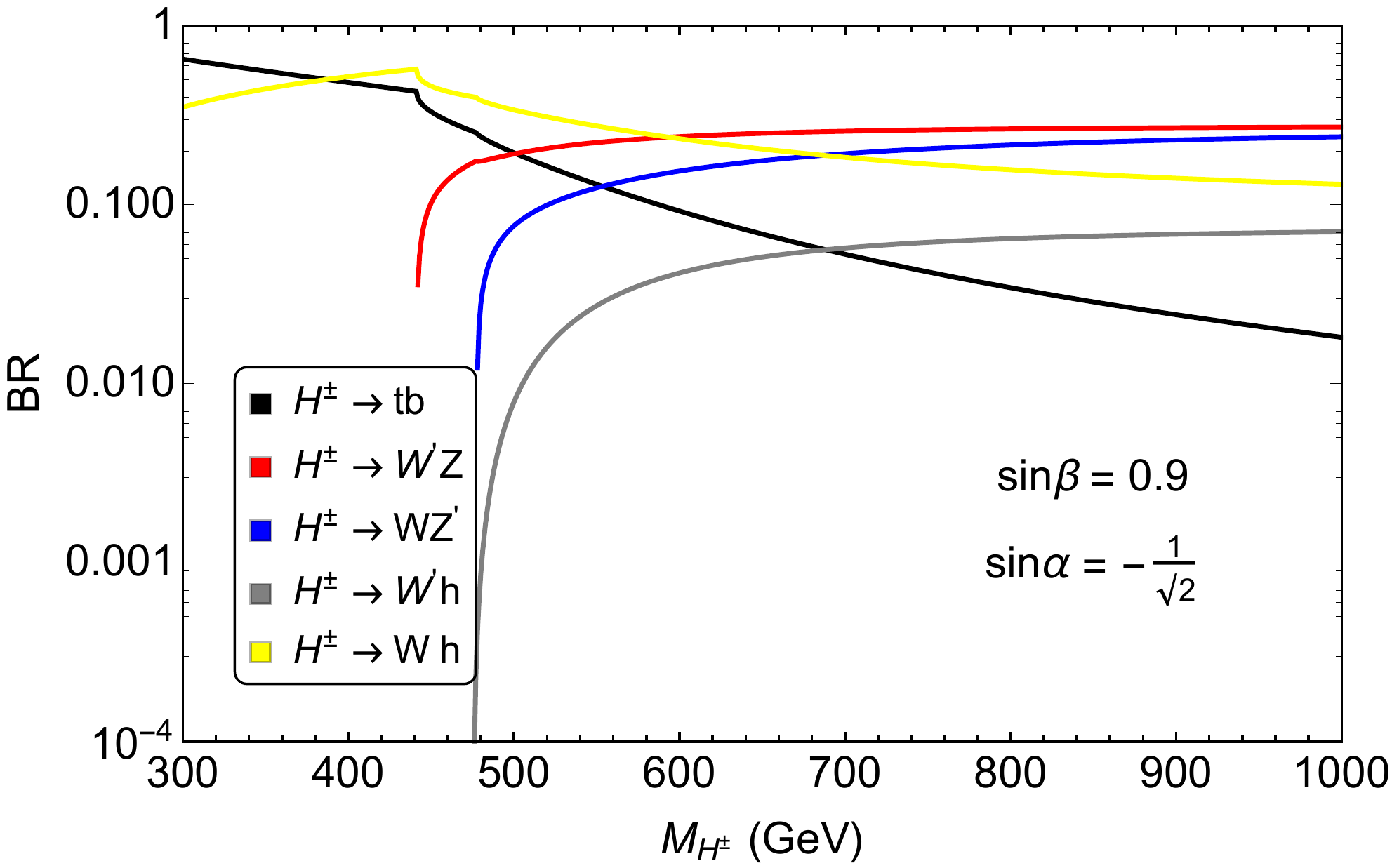}
\end{center}
\caption{The branching ratio for different charged Higgs boson decay modes for $\sin\beta$ = 0.6 and 0.9.}
\label{fig:HcBrDecay}
\end{figure}

In Fig~\ref{fig:gbHct_Sinb} we present the production cross-section for the charged Higgs boson via $p p \rightarrow H^{\pm} t$ channel for the benchmark scenario $m_{H^{\pm}}$ = 500 GeV and 700 GeV as a function of $\sin\beta$ (related to the ratio of the vevs), and in Fig.~\ref{fig:HcBrDecay}, we present the charged Higgs boson branching ratios.

\begin{figure}[h!]
\begin{center}
\includegraphics[scale=0.35]{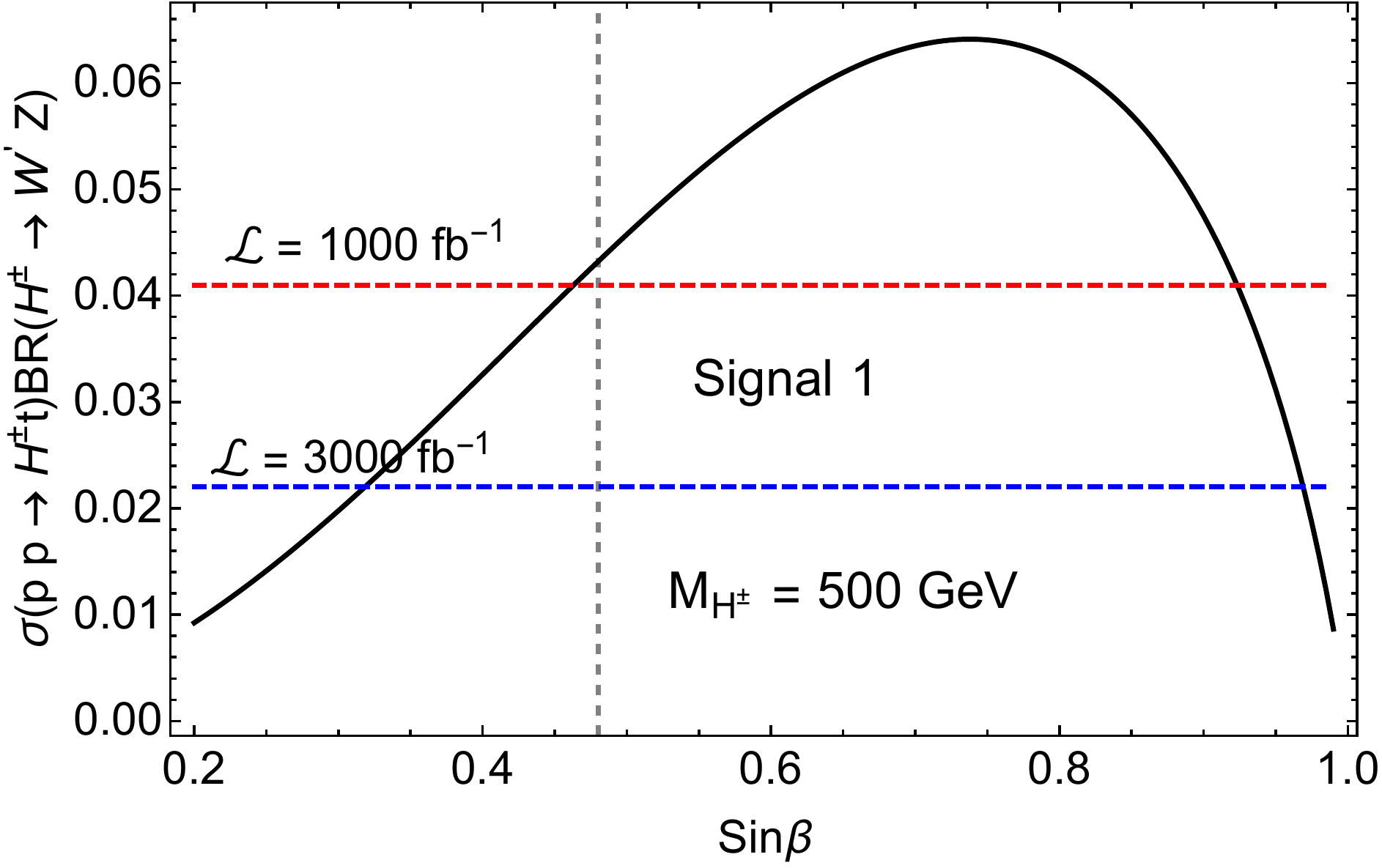}
\end{center}
\caption{The reach plots corresponding to Signal 1 for different values of $\sin\beta$ for $\sin\alpha = - \frac{1}{\sqrt{2}}$. The region left of the gray dashed vertical line is disallowed in this model from the $b\rightarrow s \gamma$ constraints.}
\label{fig:Signal1_TTM}
\end{figure}

With these numbers in place, we show in Fig.~\ref{fig:Signal1_TTM}, the reach plot in this realistic model for Signal 1 as an illustration of the methods employed in this study. It can be seen that a wide range of $\sin\beta$ allows for discovery of the charged Higgs in this channel.

\section{Conclusion}
\label{Sec:Conclusions}
While the LHC has not produced any new particle beyond the SM-like Higgs, particle physics is entering a stage where more sophisticated analysis methods need to be employed to tease out any hints of new physics. At the same time, care should be taken to exhaust all possible forms of search for possible hints of heavy gauge, scalar, or fermionic degrees of freedom already accessible at the LHC or in the near future by exploring channels and techniques not considered before. In this spirit, in this paper we have presented a case of discovering a charged Higgs boson which is traditionally looked for in $tb$ and $\tau\nu$ modes (or in $WA$ kinds of modes in 2HDM kinds of scenarios) in models with an extended gauge sector.

The extended gauge group immediately presents other, potentially new discovery modes for the charged Higgs. In this paper, we explored the decay mode $H^\pm\to W'^{\pm}Z$ and considered three different final states depending on the $W'$ decay. We have done this study under the assumption that the $W'$ is fermiophobic (and therefore can evade the direct constraints and be light) - we find, both in a toy model and in a more realistic scenario that these new signals to be very promising in terms of discovery potential. Two of the three signals that we have discussed could potentially aid discovery at 500 fb$^{-1}$, while the third might typically require higher luminosities. We close this study with the remark that models with both extended gauge and scalar sectors remain a rich source of exciting phenomenology that could be probed at the LHC in the near future.

\clearpage

\bibliography{References}

\end{document}